\documentclass[11pt]{article}
\usepackage{epsfig} 
\newcommand{\sect}[1]{ \section{#1} \setcounter{equation}{0} }

\newcommand{\half}{\mbox{\small{$\frac{1}{2}$}}}

\newcommand{\MSbar}{\overline{\mbox{MS}}} 
\newcommand{\MSbars}{\overline{\mbox{\footnotesize{MS}}}} 
\newcommand{\mMOM}{\mbox{mMOM}}
\newcommand{\mMOMs}{\mbox{\footnotesize{mMOM}}}
\newcommand{\Nc}{N_{\!c}}
\newcommand{\Nf}{N_{\!f}}
\newcommand{\NF}{N_{\!F}}
\newcommand{\NA}{N_{\!A}}

\begin{document}
\title{Renormalization group functions of QCD in the minimal MOM scheme}
\author{J.A. Gracey, \\ Theoretical Physics Division, \\ 
Department of Mathematical Sciences, \\ University of Liverpool, \\ P.O. Box 
147, \\ Liverpool, \\ L69 3BX, \\ United Kingdom.} 
\date{$\frac{}{}$} 
\maketitle 

\vspace{5cm} 
\noindent 
{\bf Abstract.} We provide the full set of renormalization group functions for
the renormalization of QCD in the minimal MOM scheme to four loops for the
colour group $SU(\Nc)$. 

\vspace{-15.5cm}
\hspace{13.5cm}
{\bf LTH 976}

\newpage

\sect{Introduction.} 

The main renormalization scheme used in quantum field theory is the modified
minimal subtraction, $\MSbar$, scheme introduced in \cite{1,2}. It has many
elegant features which can be exploited to determine the renormalization group
functions to a very high loop order. One of these is that of calculability.
Briefly, only the divergences with respect to the regulator are removed,
together with a specific finite part, $\ln(4\pi e^{-\gamma})$ where $\gamma$
is the Euler-Mascheroni constant \cite{2}. Ordinarily for conventional 
perturbation theory calculations one uses dimensional regularization in 
$d$~$=$~$4$~$-$~$2\epsilon$ dimensions and $\epsilon$ plays the role of the
regulator. Given this one need only consider the underlying massless quantum
field theory safe in the knowledge that in this renormalization scheme the
divergences will be mass independent. Thus, since massless Feynman graphs are
significantly easier to compute than massive ones, then one can extract the
ultraviolet divergences to high loop order. Moreover, using the scheme in
gauge theories with massless fields gauge symmetry is preserved, \cite{1}. Thus
$\MSbar$ has been established as the favoured scheme for many years. However, 
for certain problems it is not necessarily the best choice. For instance, in 
lattice gauge theory computations it is not practical to implement since, for 
example, it is expensive for calculating Green's functions involving 
derivatives. Instead physical schemes such as the modified regularization 
invariant (RI${}^\prime$) scheme have been introduced, \cite{3,4}. These are 
asymmetric in that the definition is related to the choice of momentum 
configuration of $3$-point functions. Related types of physical schemes, but 
not motivated by lattice considerations, are the momentum subtraction schemes 
of \cite{5} denoted by MOM. These differ from RI${}^\prime$ schemes in that the
$3$-point function momentum configuration is completely symmetric. Hence they 
do not suffer from infrared issues as the configuration does not have 
exceptional momenta. Both RI${}^\prime$ and MOM schemes differ from $\MSbar$ in
that finite pieces are absorbed into the renormalization constants which 
therefore depend on external momentum scales. As a corollary they are more 
difficult to calculate in analytically to high loop order. Irrespective of 
which scheme one chooses to use for an analysis, through the structure of the 
renormalization group equation it is possible to relate results. Thus within
perturbation theory one can compute the conversion functions which allow one to
map, for example, the coupling constant defined in one scheme to that in 
another. The other parameters and renormalization group functions can equally 
be related by the same formalism. 

A more recent development has been the introduction of another variant within
the RI${}^\prime$ and MOM family of physical renormalization schemes, \cite{6}.
It is called the minimal MOM scheme and is motivated by a property of the
ghost-gluon vertex of QCD in the Landau gauge. This property is the 
non-renormalization of the vertex, \cite{7}. However, the scheme is an
extension of the concept beyond this specific gauge in a way which preserves
a definition of the coupling constant in terms of the ghost and gluon form
factors, \cite{6}. This effective running coupling constant has been the 
subject of intense interest in recent years due to interesting features at 
medium and low energies which were noted earlier in \cite{8}. For instance, it 
is believed that there are dimension two deviations from the expected running 
when compared to pure perturbation theory. The most recent work, \cite{9}, 
appears to reaffirm this property. With the minimal MOM scheme the effective 
coupling constant is not only simple to define but from a practical point of 
view does not require full knowledge of the ghost-gluon vertex function as 
would be necessary in other schemes \cite{6}. Indeed in spite of being a 
non-exceptional momentum configuration it is numerically harder to extract a 
clean signal from the lattice for a fully symmetric vertex such as in the MOM 
context. In developing the minimal MOM scheme, \cite{6}, the four loop QCD 
$\beta$-function was determined for $SU(\Nc)$ not only in the Landau gauge but 
also for a particular formulation of a linear covariant gauge. The full set of 
renormalization group functions as far as they could be calculated were not 
given. Therefore, it is the purpose of this note to provide the wave function 
renormalization group functions to as far a loop order as is possible. For an 
arbitrary colour group this is to three loops for the wave functions and four 
loops for the $\beta$-function and quark mass anomalous dimension. Though we 
will also provide the former to four loops for $SU(\Nc)$. We will do this in 
two ways. The first is the direct evaluation of all the three loop Green's 
functions where the minimal MOM renormalization scheme definition is 
implemented directly. The second is by construction of the associated 
conversion functions and use of the renormalization group equations. This will 
serve as a check on our computations and allow us to deduce four loop
information. One of the reasons for the direct renormalization is that it 
provides a non-trivial independent check on the results of \cite{6}. There the
$\beta$-function was adduced from known finite parts of Green's functions given
in \cite{10}. As a separate exercise we choose to work in a minor variation of 
the original minimal MOM scheme and that is to renormalize the gauge parameter,
$\alpha$, in the full ethos of the minimal MOM scheme. In \cite{6} the 
renormalization of the gauge parameter is completely equivalent to that of the 
$\MSbar$ scheme. Though our results will equate for the Landau gauge. A similar
issue for $\alpha$ arises in the RI${}^\prime$ case \cite{11}. Finally, given 
the interest in the behaviour of the effective coupling constant in the Landau 
gauge and its power law deviation, we will provide the minimal MOM anomalous 
dimension of the operator thought to be associated with the dimension two 
correction which is $\half A^a_\mu A^{a\,\mu}$ where $A^a_\mu$ is the gluon. 
This is in order to allow one to perform a complete renormalization group 
running analysis in the minimal MOM scheme for such infrared problems.

The paper is organized as follows. We recall the definition and properties of
the minimal MOM scheme in section $2$ before recording our results in the
subsequent section. These include the renormalization group functions and 
the conversion functions for an arbitrary colour group. For $\MSbar$ 
renormalization group functions which are only known at four loops for 
$SU(\Nc)$, we provide the corresponding minimal MOM scheme results in section 
$4$ together with those for the dimension two operator anomalous dimension. A 
conclusion is provided in section $5$.

\sect{Formalism.}

We begin by recalling the definition of the minimal MOM scheme, \cite{6}. 
First, if we denote bare quantities in the QCD Lagrangian by the subscript
${}_{\mbox{\footnotesize{o}}}$, then in our notation the renormalization
constants, $Z_i$, are given by 
\begin{eqnarray}
A^{a \, \mu}_{\mbox{\footnotesize{o}}} &=& \sqrt{Z_A} \, A^{a \, \mu} ~~,~~
c^a_{\mbox{\footnotesize{o}}} ~=~ \sqrt{Z_c} \, c^a ~~,~~
\psi^{iI}_{\mbox{\footnotesize{o}}} ~=~ \sqrt{Z_\psi} \, \psi^{iI} \nonumber \\
\alpha_{\mbox{\footnotesize{o}}} &=& Z^{-1}_\alpha Z_A \, \alpha ~~,~~
m_{\mbox{\footnotesize{o}}} ~=~ m Z_m ~~,~~
g_{\mbox{\footnotesize{o}}} ~=~ \mu^\epsilon Z_g \, g
\label{zdef}
\end{eqnarray}
where $A^a_\mu$ is the gluon, $c^a$ is the Faddeev-Popov ghost and $\psi^i$ is
the quark. The indices have the ranges $1$~$\leq$~$a$~$\leq$~$\NA$, 
$1$~$\leq$~$i$~$\leq$~$\NF$ and $1$~$\leq$~$I$~$\leq$~$\Nf$ where $\NF$ and 
$\NA$ are the respective dimensions of the fundamental and adjoint 
representations of the colour group and $\Nf$ is the number of massive quarks 
each of the same mass $m$. The coupling constant is $g$ and $\alpha$ is the 
gauge parameter of the linear covariant gauge. The Landau gauge corresponds to 
$\alpha$~$=$~$0$. We use the above definition of the renormalization of 
$\alpha$ to be consistent with \cite{11}. Throughout we use dimensional 
regularization in $d$~$=$~$4$~$-$~$2\epsilon$ spacetime dimensions and the mass
scale $\mu$ is introduced to ensure the coupling constant is dimensionless in 
$d$-dimensions. With these formal definitions of the renormalization constants 
they are then determined explicitly by specifying a scheme to absorb the 
infinities in the various $2$ and $3$-point functions of the theory. For 
instance, $\MSbar$ corresponds to removing only the poles in $\epsilon$ 
together with a certain finite part at some subtraction point. 

For momentum subtraction schemes, denoted generally by MOM, the scheme is 
defined such that at the subtraction point the poles in $\epsilon$ together 
with all the finite part are absorbed into the renormalization constant, 
\cite{5}. For the QCD Lagrangian this produces several different schemes since 
there are several vertices which one can use to define the coupling constant 
renormalization. Choosing one, say, means that the remaining vertex functions 
are finite and consistent with the Slavnov-Taylor identities. The variation on 
this approach introduced in \cite{6} is that the $2$-point functions are 
renormalized using the MOM criterion of \cite{5} but the $3$-point vertices are
treated differently. Specifically, to ease comparison with lattice analyses the
completely symmetric subtraction point of \cite{5} is not used. Instead the 
asymmetric point is used where the external momentum of an external leg is 
nullified. Moreover, partly motivated by the non-renormalization of the 
ghost-gluon vertex in the Landau gauge, \cite{7}, the coupling constant 
renormalization is defined by ensuring that this vertex renormalization 
constant is the {\em same} as the $\MSbar$ one. One benefit of this, \cite{6}, 
is that to define the scheme one only needs to know the vertex structure in the
$\MSbar$ scheme which reduces work for non-perturbative applications. In our 
notation, (\ref{zdef}), this corresponds to, \cite{6},
\begin{equation} 
Z_g^{\MSbars} \sqrt{Z_A^{\MSbars}} Z_c^{\MSbars} ~=~
Z_g^{\mMOMs} \sqrt{Z_A^{\mMOMs}} Z_c^{\mMOMs} 
\label{zgmm}
\end{equation} 
where $\mMOM$ denotes the minimal MOM scheme. Though, in this formal definition 
it is important to appreciate that the variables $g$ and $\alpha$ on either 
side of the equation are in different schemes. We note that throughout our 
convention is that when a scheme is specified as a label on a quantity then it 
is a function of the parameters $g$ and $\alpha$ in that scheme. With 
(\ref{zgmm}) then all the renormalization constants of massless QCD are defined
for the minimal MOM scheme, \cite{6}. As noted earlier in \cite{6} the gauge 
parameter renormalization was treated as an $\MSbar$ one rather than define it 
as the full MOM renormalization as used in \cite{5}. Therefore, we will follow 
the approach of \cite{5} here and have a minimal MOM $\alpha$. From the 
practical point of view our results in the Landau gauge will be the same and 
differ only in the $\alpha$ dependent part.
 
The procedure we have used is to apply the {\sc Mincer} algorithm,
\cite{12}, to the massless QCD Lagrangian and compute all the $2$-point 
functions as well as the ghost-gluon vertex at the asymmetric point. The quark
mass anomalous dimension will be discussed later. This algorithm evaluates 
massless three loop $2$-point functions to the finite part in dimensional 
regularization. It has been encoded, \cite{13}, in the symbolic manipulation 
language {\sc Form}, \cite{14}, which is our main computational tool. The 
Feynman diagrams are generated by {\sc Qgraf}, \cite{15}, and the output 
converted into {\sc Form} input notation. As all graphs are evaluated to the 
finite parts then we can extract the explicit renormalization constants in the
minimal MOM. We do this first by renormalizing the $2$-point functions before 
defining the coupling constant renormalization via (\ref{zgmm}). Then one
proceeds to the next loop order. In using the definition of the coupling
constant renormalization we have to relate the parameters between the schemes.
For the gauge parameter this is given by 
\begin{equation}
\alpha_{\mMOMs}(\mu) ~=~ \frac{Z_A^{\MSbars}}{Z_A^{\mMOMs}} \,
\alpha_{\MSbars}(\mu) 
\end{equation}
where we used the fact that we are in a linear covariant gauge which implies
$Z_\alpha$~$=$~$1$. To ensure a finite expression in $\epsilon$ emerges the 
parameters within $Z_A^{\mMOMs}$ have to be converted to their $\MSbar$ 
partners. This is achieved order by order in perturbation theory. We have 
determined these to three loops and, with $a$~$=$~$g^2/(16\pi^2)$, found
\begin{eqnarray}
a^{\mMOMs} &=& a + \left[ 9 \alpha^2 C_A + 18 \alpha C_A + 169 C_A 
- 80 \Nf T_F \right] \frac{a^2}{36} \nonumber \\
&& + \left[ 405 \alpha^3 C_A^2 - 486 \zeta(3) \alpha^2 C_A^2 
+ 2835 \alpha^2 C_A^2 + 3564 \zeta(3) \alpha C_A^2 + 2421 \alpha C_A^2 
\right. \nonumber \\
&& \left. ~~~
- 1440 \alpha C_A \Nf T_F - 6318 \zeta(3) C_A^2 + 76063 C_A^2 
- 10368 \zeta(3) C_A \Nf T_F 
\right. \nonumber \\
&& \left. ~~~
- 50656 C_A \Nf T_F 
+ 20736 \zeta(3) C_F \Nf T_F 
- 23760 C_F \Nf T_F + 6400 \Nf^2 T_F^2 \right] \frac{a^3}{1296} \nonumber \\
&& + \left[ - 10692 \zeta(3) \alpha^4 C_A^3 + 8505 \zeta(5) \alpha^4 C_A^3 
+ 41067 \alpha^4 C_A^3 - 59292 \zeta(3) \alpha^3 C_A^3 
\right. \nonumber \\
&& \left. ~~~
- 4860 \zeta(5) \alpha^3 C_A^3 + 293301 \alpha^3 C_A^3 
- 138024 \zeta(3) \alpha^2 C_A^3 - 85050 \zeta(5) \alpha^2 C_A^3 
\right. \nonumber \\
&& \left. ~~~
+ 1315035 \alpha^2 C_A^3 
+ 46656 \zeta(3) \alpha^2 C_A^2 \Nf T_F 
- 322056 \alpha^2 C_A^2 \Nf T_F
\right. \nonumber \\
&& \left. ~~~
+ 3355020 \zeta(3) \alpha C_A^3 
- 860220 \zeta(5) \alpha C_A^3 + 1277496 \alpha C_A^3 
\right. \nonumber \\
&& \left. ~~~
- 2115072 \zeta(3) \alpha C_A^2 \Nf T_F - 581760 \alpha C_A^2 \Nf T_F 
+ 373248 \zeta(3) \alpha C_A C_F \Nf T_F 
\right. \nonumber \\
&& \left. ~~~
- 427680 \alpha C_A C_F \Nf T_F 
+ 331776 \zeta(3) \alpha C_A \Nf^2 T_F^2 + 32256 \alpha C_A \Nf^2 T_F^2 
\right. \nonumber \\
&& \left. ~~~
- 6552900 \zeta(3) C_A^3 - 1896615 \zeta(5) C_A^3 
+ 42074947 C_A^3 
- 4499712 \zeta(3) C_A^2 \Nf T_F 
\right. \nonumber \\
&& \left. ~~~
+ 2488320 \zeta(5) C_A^2 \Nf T_F 
- 38975424 C_A^2 \Nf T_F + 15303168 \zeta(3) C_A C_F \Nf T_F 
\right. \nonumber \\
&& \left. ~~~
+ 3732480 \zeta(5) C_A C_F \Nf T_F 
- 23755968 C_A C_F \Nf T_F 
+ 3068928 \zeta(3) C_A \Nf^2 T_F^2 
\right. \nonumber \\
&& \left. ~~~
+ 9209280 C_A \Nf^2 T_F^2 
+ 4603392 C_F^2 \Nf T_F - 7464960 \zeta(5) C_F^2 \Nf T_f T_F 
\right. \nonumber \\
&& \left. ~~~
+ 1482624 C_F^2 \Nf T_F 
- 6469632 \zeta(3) C_F \Nf^2 T_F^2 
\right. \nonumber \\
&& \left. ~~~
+ 8065152 C_F \Nf^2 T_F^2 - 512000 \Nf^3 T_F^3 \right] 
\frac{a^4}{46656} ~+~ O(a^5)
\label{amap}
\end{eqnarray}
and
\begin{eqnarray}
\alpha^{\mMOMs} &=& 
\alpha + \left[ - 9 \alpha^2 C_A - 18 \alpha C_A - 97 C_A + 80 \Nf T_F
\right] \frac{\alpha a}{36} \nonumber \\
&& +~ \left[ 18 \alpha^4 C_A^2 - 18 \alpha^3 C_A^2 + 190 \alpha^2 C_A^2 
- 320 \alpha^2 C_A \Nf T_F - 576 \zeta(3) \alpha C_A^2 + 463 \alpha C_A^2 
\right. \nonumber \\
&& \left. ~~~~
- 320 \alpha C_A \Nf T_F + 864 \zeta(3) C_A^2 - 7143 C_A^2 
+ 2304 \zeta(3) C_A \Nf T_F + 4248 C_A \Nf T_F 
\right. \nonumber \\
&& \left. ~~~~
- 4608 \zeta(3) C_F \Nf T_F 
+ 5280 C_F \Nf T_F \right] \frac{\alpha a^2}{288} \nonumber \\
&& +~ \left[ - 486 \alpha^6 C_A^3 + 1944 \alpha^5 C_A^3 
+ 4212 \zeta(3) \alpha^4 C_A^3 - 5670 \zeta(5) \alpha^4 C_A^3 
- 18792 \alpha^4 C_A^3 
\right. \nonumber \\
&& \left. ~~~~
+ 12960 \alpha^4 C_A^2 \Nf T_F 
+ 48276 \zeta(3) \alpha^3 C_A^3 - 6480 \zeta(5) \alpha^3 C_A^3 
- 75951 \alpha^3 C_A^3 
\right. \nonumber \\
&& \left. ~~~~
- 8640 \alpha^3 C_A^2 \Nf T_F 
- 52164 \zeta(3) \alpha^2 C_A^3 + 2916 \zeta(4) \alpha^2 C_A^3 
+ 124740 \zeta(5) \alpha^2 C_A^3 
\right. \nonumber \\
&& \left. ~~~~
+ 92505 \alpha^2 C_A^3 
- 129600 \zeta(3) \alpha^2 C_A^2 \Nf T_F - 147288 \alpha^2 C_A^2 \Nf T_F 
\right. \nonumber \\
&& \left. ~~~~
+ 248832 \zeta(3) \alpha^2 C_A C_F \Nf T_F - 285120 \alpha^2 C_A C_F \Nf T_F 
- 38400 \alpha^2 C_A \Nf^2 T_F^2 
\right. \nonumber \\
&& \left. ~~~~
- 1303668 \zeta(3) \alpha C_A^3 
+ 11664 \zeta(4) \alpha C_A^3 + 447120 \zeta(5) \alpha C_A^3 
+ 354807 \alpha C_A^3 
\right. \nonumber \\
&& \left. ~~~~
+ 698112 \zeta(3) \alpha C_A^2 \Nf T_F 
- 312336 \alpha C_A^2 \Nf T_F 
+ 248832 \zeta(3) \alpha C_A C_F \Nf T_F 
\right. \nonumber \\
&& \left. ~~~~
- 285120 \alpha C_A C_F \Nf T_F - 221184 \zeta(3) \alpha C_A \Nf^2 T_F^2 
+ 55296 \alpha C_A \Nf^2 T_F^2 
\right. \nonumber \\
&& \left. ~~~~
+ 2007504 \zeta(3) C_A^3 + 8748 \zeta(4) C_A^3 
+ 1138050 \zeta(5) C_A^3 
- 10221367 C_A^3 
\right. \nonumber \\
&& \left. ~~~~
+ 1505088 \zeta(3) C_A^2 \Nf T_F 
- 279936 \zeta(4) C_A^2 \Nf T_F - 1658880 \zeta(5) C_A^2 \Nf T_F 
\right. \nonumber \\
&& \left. ~~~~
+ 9236488 C_A^2 \Nf T_F - 5156352 \zeta(3) C_A C_F \Nf T_F 
+ 373248 \zeta(4) C_A C_F \Nf T_F 
\right. \nonumber \\
&& \left. ~~~~
- 2488320 \zeta(5) C_A C_F \Nf T_F 
+ 9293664 C_A C_F \Nf T_F - 884736 \zeta(3) C_A \Nf^2 T_F^2 
\right. \nonumber \\
&& \left. ~~~~
- 1343872 C_A \Nf^2 T_F^2 - 3068928 \zeta(3) C_F^2 \Nf T_F 
+ 4976640 \zeta(5) C_F^2 \Nf T_F 
\right. \nonumber \\
&& \left. ~~~~
- 988416 C_F^2 \Nf T_F 
+ 2101248 \zeta(3) C_F \Nf^2 T_F^2 - 2842368 C_F \Nf^2 T_F^2 \right]
\frac{\alpha a^3}{31104} \nonumber \\
&& +~ O(a^4) 
\label{almap}
\end{eqnarray}
where $\zeta(z)$ is the Riemann zeta function. The group Casimirs are defined 
by
\begin{equation}
\mbox{Tr} \left( T^a T^b \right) ~=~ T_F \delta^{ab} ~~,~~
T^a T^a ~=~ C_F I ~~,~~
f^{acd} f^{bcd} ~=~ C_A \delta^{ab}
\end{equation}
where $T^a$ are the generators of the colour group whose structure functions
are $f^{abc}$. In (\ref{amap}) and (\ref{almap}) the variables on the right
hand side are in the $\MSbar$ scheme. For the Landau gauge it is easy to see 
that then the parameters coincide. We have checked that (\ref{amap}) agrees
with the alternative definition of the mapping given in \cite{6} based on the 
actual finite parts of the gluon and ghost $2$-point functions after their
$\MSbar$ renormalization. 

While we will perform a direct evaluation of the renormalization constants in
the minimal MOM, there are several checks which will be carried out. One is to
exploit properties of the renormalization group equation which allows one to
map the anomalous dimensions deduced in each scheme via conversion functions
which are denoted by $C_i(a,\alpha)$ where $i$ will be a label corresponding
to a field or a parameter. First, we will perform the explicit renormalization
in the minimal MOM and deduce the anomalous dimensions directly. Then we will
compute the conversion functions and from these construct the anomalous
dimensions indirectly. Thus if we define the conversion functions by 
\begin{equation}
C^{\mMOMs}_g(a,\alpha) ~=~ \frac{Z_g^{\mMOMs}}{Z_g^{\MSbars}} ~~~~,~~~~
C^{\mMOMs}_\phi(a,\alpha) ~=~ \frac{Z_\phi^{\mMOMs}}{Z_\phi^{\MSbars}}
\end{equation}
where $\phi$~$\in$~$\{A,c,\psi\}$, then the minimal MOM renormalization group 
functions are given by 
\begin{eqnarray}
\beta^{\mbox{$\mMOMs$}} ( a_{\mbox{$\mMOMs$}}, \alpha_{\mbox{$\mMOMs$}} ) &=&
\left[ \beta^{\mbox{$\MSbars$}}( a_{\mbox{$\MSbars$}} )
\frac{\partial a_{\mbox{$\mMOMs$}}}{\partial a_{\mbox{$\MSbars$}}} \right. 
\nonumber \\
&& \left. ~+~ \alpha_{\mbox{$\MSbars$}} \gamma^{\mbox{$\MSbars$}}_\alpha
( a_{\mbox{$\MSbars$}}, \alpha_{\mbox{\footnotesize{$\MSbars$}}} )
\frac{\partial a_{\mbox{$\mMOMs$}}}{\partial \alpha_{\mbox{$\MSbars$}}}
\right]_{ \MSbars \rightarrow \mMOMs }
\end{eqnarray}
and
\begin{eqnarray}
\gamma_\phi^{\mMOMs} ( a_{\mMOMs}, \alpha_{\mMOMs} )
\!\!&=& \!\!\!\!\! \left[ \gamma_\phi^{\MSbars} \left( a_{\MSbars}, 
\alpha_{\MSbars} \right) \right. \nonumber \\
&& \!\!\!\!\! \left. +\, \beta^{\MSbars}\left(a_{\MSbars}\right)
\frac{\partial ~}{\partial a_{\MSbars}} \ln C_\phi^{\mMOMs}
\left(a_{\MSbars},\alpha_{\MSbars}\right) \right. \nonumber \\
&& \!\!\!\!\! \left. +\, \alpha_{\MSbars} \gamma^{\MSbars}_\alpha \!
\left(a_{\MSbars},\alpha_{\MSbars}\right) \!\!
\frac{\partial ~}{\partial \alpha_{\MSbars}}
\ln C_\phi^{\mMOMs} \!\! \left(a_{\MSbars},\alpha_{\MSbars}\right) \!
\right]_{ \MSbars \rightarrow \mMOMs } \!\!\!\! . \nonumber \\
\end{eqnarray}
Here $\MSbar$~$\rightarrow$~$\mMOM$ means that after computing the right hand
side the expression will be a function of $\MSbar$ variables and these must
therefore be converted to minimal MOM ones. The relations are given by 
inverting (\ref{amap}) and (\ref{almap}). One benefit of this formalism is
that it can be exploited to produce the {\em four} loop anomalous dimensions
and $\beta$-function. The reason for this is that the three loop conversion 
functions give a four loop contribution to the minimal MOM anomalous
dimensions and $\beta$-function and as the $\MSbar$ versions of these are 
known, \cite{16,17,18,19,20,21,22,23,24,25,26}, then the left hand side can be 
deduced at four loops.

\sect{Results.}

We now formally record our results. The one loop expressions will be the same
as the $\MSbar$ ones since that term is scheme independent. This includes the
$\beta$-function as we are using a mass dependent renormalization scheme and
only in mass independent schemes is the two loop term scheme independent.
Moreover, the $\beta$-function will be gauge dependent for the same reason.
Therefore, we have\footnote{A data file is attached which gives an electronic
version of all our expressions.} 
\begin{eqnarray}
\beta^{\mbox{$\mMOMs$}}(a,\alpha) &=& -~ \left[ 11 C_A - 4 \Nf T_F \right] 
\frac{a^2}{3} \nonumber \\
&& +~ \left[ -~ 3 \alpha^3 C_A^2 + 10 \alpha^2 C_A^2 - 8 \alpha^2 C_A \Nf T_F 
+ 13 \alpha C_A^2 - 8 \alpha C_A \Nf T_F - 136 C_A^2 \right. \nonumber \\
&& \left. ~~~~+ 
80 C_A \Nf T_F + 48 C_F \Nf T_F \right] \frac{a^3}{12} \nonumber \\
&& +~ \left[ - 165 \alpha^4 C_A^3 + 24 \alpha^4 C_A^2 \Nf T_F 
+ 108 \zeta(3) \alpha^3 C_A^3 - 189 \alpha^3 C_A^3 
\right. \nonumber \\
&& \left. ~~~~
- 144 \alpha^3 C_A^2 \Nf T_F
- 468 \zeta(3) \alpha^2 C_A^3 + 2175 \alpha^2 C_A^3 
+ 144 \zeta(3) \alpha^2 C_A^2 \Nf T_F 
\right. \nonumber \\
&& \left. ~~~~
- 1656 \alpha^2 C_A^2 \Nf T_F - 864 \alpha^2 C_A C_F \Nf T_F 
- 1188 \zeta(3) \alpha C_A^3 + 3291 \alpha C_A^3 
\right. \nonumber \\
&& \left. ~~~~
- 1776 \alpha C_A^2 \Nf T_F - 1152 \alpha C_A C_F \Nf T_F 
+ 5148 \zeta(3) C_A^3 - 38620 C_A^3 
\right. \nonumber \\
&& \left. ~~~~
+ 6576 \zeta(3) C_A^2 \Nf T_F + 32144 C_A^2 \Nf T_F 
- 16896 \zeta(3) C_A C_F \Nf T_F 
\right. \nonumber \\
&& \left. ~~~~
+ 20512 C_A C_F \Nf T_F 
- 3072 \zeta(3) C_A \Nf^2 T_F^2 - 4416 C_A \Nf^2 T_F^2 
\right. \nonumber \\
&& \left. ~~~~
- 576 C_F^2 \Nf T_F 
+ 6144 \zeta(3) C_F \Nf^2 T_F^2 - 5888 C_F \Nf^2 T_F^2 \right] \frac{a^4}{288}
\nonumber \\  
&& + \left[ 864 \zeta(3) \alpha^5 C_A^4 - 3780 \zeta(5) \alpha^5 C_A^4 
- 11745 \alpha^5 C_A^4 + 1728 \alpha^5 C_A^3 \Nf T_F 
\right. \nonumber \\
&& \left. ~~~~
+ 32472 \zeta(3) \alpha^4 C_A^4 + 4140 \zeta(5) \alpha^4 C_A^4 
- 81549 \alpha^4 C_A^4 
\right. \nonumber \\
&& \left. ~~~~
- 1440 \zeta(3) \alpha^4 C_A^3 \Nf T_F 
- 5040 \zeta(5) \alpha^4 C_A^3 \Nf T_F + 7200 \alpha^4 C_A^3 \Nf T_F 
\right. \nonumber \\
&& \left. ~~~~
+ 7776 \alpha^4 C_A^2 C_F \Nf T_F 
+ 47052 \zeta(3) \alpha^3 C_A^4 
+ 19800 \zeta(5) \alpha^3 C_A^4 
\right. \nonumber \\
&& \left. ~~~~
- 81873 \alpha^3 C_A^4 
+ 18432 \zeta(3) \alpha^3 C_A^3 \Nf T_F 
+ 1440 \zeta(5) \alpha^3 C_A^3 \Nf T_F 
\right. \nonumber \\
&& \left. ~~~~
- 67752 \alpha^3 C_A^3 \Nf T_F 
- 7776 \alpha^3 C_A^2 C_F \Nf T_F 
- 397368 \zeta(3) \alpha^2 C_A^4 
\right. \nonumber \\
&& \left. ~~~~
+ 152280 \zeta(5) \alpha^2 C_A^4 
+ 1028898 \alpha^2 C_A^4 
- 36576 \zeta(3) \alpha^2 C_A^3 \Nf T_F 
\right. \nonumber \\
&& \left. ~~~~
- 1098936 \alpha^2 C_A^3 \Nf T_F 
+ 639360 \zeta(3) \alpha^2 C_A^2 C_F \Nf T_F 
\right. \nonumber \\
&& \left. ~~~~
- 790272 \alpha^2 C_A^2 C_F \Nf T_F 
+ 73728 \zeta(3) \alpha^2 C_A^2 \Nf^2 T_F^2 
+ 133632 \alpha^2 C_A^2 \Nf^2 T_F^2 
\right. \nonumber \\
&& \left. ~~~~
+ 20736 \alpha^2 C_A C_F^2 \Nf T_F 
- 221184 \zeta(3) \alpha^2 C_A C_F \Nf^2 T_F^2 
\right. \nonumber \\
&& \left. ~~~~
+ 211968 \alpha^2 C_A C_F \Nf^2 T_F^2 - 2400708 \zeta(3) \alpha C_A^4 
+ 987660 \zeta(5) \alpha C_A^4 
\right. \nonumber \\
&& \left. ~~~~
+ 1719423 \alpha C_A^4 
+ 1655712 \zeta(3) \alpha C_A^3 \Nf T_F 
- 254880 \zeta(5) \alpha C_A^3 \Nf T_F 
\right. \nonumber \\
&& \left. ~~~~
- 1817880 \alpha C_A^3 \Nf T_F 
+ 798336 \zeta(3) \alpha C_A^2 C_F \Nf T_F 
\right. \nonumber \\
&& \left. ~~~~
- 1030752 \alpha C_A^2 C_F \Nf T_F 
- 617472 \zeta(3) \alpha C_A^2 \Nf^2 T_F^2 
+ 391488 \alpha C_A^2 \Nf^2 T_F^2 
\right. \nonumber \\
&& \left. ~~~~
+ 31104 \alpha C_A C_F^2 \Nf T_F 
- 331776 \zeta(3) \alpha C_A C_F \Nf^2 T_F^2 
\right. \nonumber \\
&& \left. ~~~~
+ 317952 \alpha C_A C_F \Nf^2 T_F^2 
+ 98304 \zeta(3) \alpha C_A \Nf^3 T_F^3 - 24576 \alpha C_A \Nf^3 T_F^3 
\right. \nonumber \\
&& \left. ~~~~
+ 5509416 \zeta(3) C_A^4 
+ 3090780 \zeta(5) C_A^4 - 22106704 C_A^4 
\right. \nonumber \\
&& \left. ~~~~
- 1217376 \zeta(3) C_A^3 \Nf T_F 
- 5178960 \zeta(5) C_A^3 \Nf T_F 
+ 23501280 C_A^3 \Nf T_F 
\right. \nonumber \\
&& \left. ~~~~
- 7050240 \zeta(3) C_A^2 C_F \Nf T_F 
- 6082560 \zeta(5) C_A^2 C_F \Nf T_F 
\right. \nonumber \\
&& \left. ~~~~
+ 17477280 C_A^2 C_F \Nf T_F 
- 1654272 \zeta(3) C_A^2 \Nf^2 T_F^2 
\right. \nonumber \\
&& \left. ~~~~
+ 1474560 \zeta(5) C_A^2 \Nf^2 T_F^2 - 5719680 C_A^2 \Nf^2 T_F^2 
\right. \nonumber \\
&& \left. ~~~~
- 7907328 \zeta(3) C_A C_F^2 \Nf T_F 
+ 12165120 \zeta(5) C_A C_F^2 \Nf T_F 
\right. \nonumber \\
&& \left. ~~~~
- 607104 C_A C_F^2 \Nf T_F 
+ 4755456 \zeta(3) C_A C_F \Nf^2 T_F^2 
\right. \nonumber \\
&& \left. ~~~~
+ 2211840 \zeta(5) C_A C_F \Nf^2 T_F^2 - 10861056 C_A C_F \Nf^2 T_F^2 
\right. \nonumber \\
&& \left. ~~~~
+ 344064 \zeta(3) C_A \Nf^3 T_F^3 + 229376 C_A \Nf^3 T_F^3 
- 476928 C_F^3 \Nf T_F 
\right. \nonumber \\
&& \left. ~~~~
+ 3538944 \zeta(3) C_F^2 \Nf^2 T_F^2 
- 4423680 \zeta(5) C_F^2 \Nf^2 T_F^2 + 267264 C_F^2 \Nf^2 T_F^2 
\right. \nonumber \\
&& \left. ~~~~
- 884736 \zeta(3) C_F \Nf^3 T_F^3 + 1327104 C_F \Nf^3 T_F^3 
- 2433024 \zeta(3) \frac{d^{(4)}_{AA}}{\NA} 
\right. \nonumber \\
&& \left. ~~~~
+ 92160 \frac{d^{(4)}_{AA}}{\NA} 
+ 5750784 \zeta(3) \frac{d^{(4)}_{FA}}{\NA} \Nf 
- 589824 \frac{d^{(4)}_{FA}}{\NA} \Nf 
\right. \nonumber \\
&& \left. ~~~~
- 1769472 \zeta(3) \frac{d^{(4)}_{FF}}{\NA} \Nf^2 
+ 811008 \frac{d^{(4)}_{FF}}{\NA} \Nf^2 
\right] \frac{a^5}{10368} ~+~ O(a^6) ~.
\end{eqnarray}
As the four loop $\MSbar$ $\beta$-function was computed for an arbitrary colour
group, \cite{23}, the general colour group Casimirs appear. In our notation 
they are defined by
\begin{equation}
d^{(4)}_{FF} ~=~ d_F^{abcd} d_F^{abcd} ~~,~~
d^{(4)}_{FA} ~=~ d_F^{abcd} d_A^{abcd} ~~,~~
d^{(4)}_{AA} ~=~ d_A^{abcd} d_A^{abcd} 
\end{equation}
and the totally symmetric rank $4$ colour tensor is defined by, \cite{27}, 
\begin{equation}
d_R^{abcd} ~=~  \frac{1}{6} \mbox{Tr} \left( T^a T^{(b} T^c T^{d)} \right)
\end{equation}
where the group generators are in the $R$ representation.  

For the anomalous dimensions only the three loop $\MSbar$ expressions are known
for an arbitrary colour group, \cite{21}. Thus to the same order the minimal 
MOM expressions are 
\begin{eqnarray}
\gamma_A^{\mbox{$\mMOMs$}}(a,\alpha) &=& 
\left[ 3 \alpha C_A - 13 C_A + 8 \Nf T_F \right] \frac{a}{6} \nonumber \\
&& + \left[ - 6 \alpha^3 C_A^2 + 17 \alpha^2 C_A^2 - 16 \alpha^2 C_A \Nf T_F 
+ 17 \alpha C_A^2 - 16 \alpha C_A \Nf T_F - 170 C_A^2 
\right. \nonumber \\
&& \left. ~~~
+ 136 C_A \Nf T_F + 96 C_F \Nf T_F \right] \frac{a^2}{24} \nonumber \\
&& + \left[ - 165 \alpha^4 C_A^3 + 24 \alpha^4 C_A^2 \Nf T_F 
+ 54 \zeta(3) \alpha^3 C_A^3 - 126 \alpha^3 C_A^3 - 144 \alpha^3 C_A^2 \Nf T_F 
\right. \nonumber \\
&& \left. ~~~
- 576 \zeta(3) \alpha^2 C_A^3 + 1761 \alpha^2 C_A^3 
+ 144 \zeta(3) \alpha^2 C_A^2 \Nf T_F - 1512 \alpha^2 C_A^2 \Nf T_F 
\right. \nonumber \\
&& \left. ~~~
- 864 \alpha^2 C_A C_F \Nf T_F - 774 \zeta(3) \alpha C_A^3 + 102 \alpha C_A^3
- 288 \zeta(3) \alpha C_A^2 \Nf T_F 
\right. \nonumber \\
&& \left. ~~~
- 600 \alpha C_A^2 \Nf T_F 
- 1152 \alpha C_A C_F \Nf T_F + 3456 \zeta(3) C_A^3 - 23032 C_A^3 
\right. \nonumber \\
&& \left. ~~~
+ 6288 \zeta(3) C_A^2 \Nf T_F + 21320 C_A^2 \Nf T_F 
- 16896 \zeta(3)C_A C_F \Nf T_F 
\right. \nonumber \\
&& \left. ~~~
+ 19648 C_A C_F \Nf T_F - 3072 \zeta(3) C_A \Nf^2 T_F^2 - 2496 C_A \Nf^2 T_F^2 
\right. \nonumber \\
&& \left. ~~~
- 576 C_F^2 \Nf T_F 
+ 6144 \zeta(3) C_F \Nf^2 T_F^2 - 5888 C_F \Nf^2 T_F^2 \right] 
\frac{a^3}{288} ~+~ O(a^4) \nonumber \\
\gamma_c^{\mbox{$\mMOMs$}}(a,\alpha) &=& 
\left[ \alpha - 3 \right] \frac{C_A a}{4} 
+ \left[3 \alpha^2 C_A - 3 \alpha C_A - 34 C_A + 8 \Nf T_F \right]
\frac{C_A a^2}{16} \nonumber \\
&& + \left[ 54 \zeta(3) \alpha^3 C_A^2 - 45 \alpha^3 C_A^2 
- 36 \zeta(3) \alpha^2 C_A^2 + 216 \alpha^2 C_A^2 - 48 \alpha^2 C_A \Nf T_F 
\right. \nonumber \\
&& \left. ~~~
+ 42 \zeta(3) \alpha C_A^2 + 109 \alpha C_A^2 + 96 \zeta(3) \alpha C_A \Nf T_F 
- 152 \alpha C_A \Nf T_F 
\right. \nonumber \\
&& \left. ~~~
+ 564 \zeta(3) C_A^2 
- 5196 C_A^2 + 96 \zeta(3) C_A \Nf T_F + 3608 C_A \Nf T_F + 288 C_F \Nf T_F 
\right. \nonumber \\
&& \left. ~~~
- 640 \Nf^2 T_F^2 \right] \frac{C_A a^3}{192} ~+~ O(a^4) \nonumber \\ 
\gamma_\psi^{\mbox{$\mMOMs$}}(a,\alpha) &=& \alpha C_F a 
+ C_F \left[ 3 \alpha^2 C_A + 6 \alpha C_A + 25 C_A - 6 C_F - 8 \Nf T_F \right]
\frac{C_Fa^2}{4} \nonumber \\
&& +~ \left[ 18 \zeta(3) \alpha^3 C_A^2 + 27 \alpha^3 C_A^2 
- 24 \alpha^3 C_A C_F - 90 \zeta(3) \alpha^2 C_A^2 + 123 \alpha^2 C_A^2 
\right. \nonumber \\
&& \left. ~~~~
-~ 36 \alpha^2 C_A C_F + 48 \alpha^2 C_A \Nf T_F - 618 \zeta(3) \alpha C_A^2 
+ 395 \alpha C_A^2 + 72 \alpha C_A C_F
\right. \nonumber \\
&& \left. ~~~~
+~ 192 \zeta(3) \alpha C_A \Nf T_F - 64 \alpha C_A \Nf T_F 
- 1470 \zeta(3)C_A^2 + 3843 C_A^2
\right. \nonumber \\
&& \left. ~~~~
+~ 576 \zeta(3) C_A C_F - 1260 C_A C_F + 384 \zeta(3) C_A \Nf T_F 
- 1840 C_A \Nf T_F 
\right. \nonumber \\
&& \left. ~~~~
+ 72 C_F^2 - 96 C_F \Nf T_F + 128 \Nf^2 T_F^2 \right] 
\frac{C_F a^3}{48} ~+~ O(a^4) ~. 
\end{eqnarray} 
We have checked explicitly that the gauge parameter satisfies 
\begin{equation} 
\gamma_\alpha^{\mbox{$\mMOMs$}}(a,\alpha) ~=~ -~ 
\gamma_A^{\mbox{$\mMOMs$}}(a,\alpha) 
\end{equation} 
which is a check on our calculation.

Having provided the anomalous dimensions we have checked that they are 
completely reproduced using the conversion function approach. The explicit
forms of these functions are 
\begin{eqnarray}
C_g(a,\alpha) &=& 
1 + \left[ - 9 \alpha^2 C_A - 18 \alpha C_A - 169 C_A + 80 \Nf T_F \right]
\frac{a}{72} \nonumber \\
&& + \left[ 243 \alpha^4 C_A^2 - 648 \alpha^3 C_A^2
+ 1944 \zeta(3) \alpha^2 C_A^2 - 1242 \alpha^2 C_A^2
- 4320 \alpha^2 C_A \Nf T_F 
\right. \nonumber \\
&& \left. ~~~
- 14256 \zeta(3) \alpha C_A^2 + 8568 \alpha C_A^2
- 2880 \alpha C_A \Nf T_F + 25272 \zeta(3) C_A^2 - 218569 C_A^2
\right. \nonumber \\
&& \left. ~~~
+ 41472 \zeta(3) C_A \Nf T_F + 121504 C_A \Nf T_F - 82944 \zeta(3) C_F \Nf T_F
+ 95040 C_F \Nf T_F
\right. \nonumber \\
&& \left. ~~~
- 6400 \Nf^2 T_F^2 \right] \frac{a^2}{10368}
\nonumber \\
&& + \left[ - 3645 \alpha^6 C_A^3 + 21870 \alpha^5 C_A^3
+ 33048 \zeta(3) \alpha^4 C_A^3 - 68040 \zeta(5) \alpha^4 C_A^3
\right. \nonumber \\
&& \left. ~~~
- 183951 \alpha^4 C_A^3 
+ 97200 \alpha^4 C_A^2 \Nf T_F
+ 754272 \zeta(3) \alpha^3 C_A^3 + 38880 \zeta(5) \alpha^3 C_A^3
\right. \nonumber \\
&& \left. ~~~
- 1501740 \alpha^3 C_A^3 
- 155520 \alpha^3 C_A^2 \Nf T_F
+ 206064 \zeta(3) \alpha^2 C_A^3 + 680400 \zeta(5) \alpha^2 C_A^3
\right. \nonumber \\
&& \left. ~~~
- 710235 \alpha^2 C_A^3 
- 1026432 \zeta(3) \alpha^2 C_A^2 \Nf T_F
- 1887840 \alpha^2 C_A^2 \Nf T_F 
\right. \nonumber \\
&& \left. ~~~
+ 2239488 \zeta(3) \alpha^2 C_A C_F \Nf T_F
- 2566080 \alpha^2 C_A C_F \Nf T_F 
- 172800 \alpha^2 C_A \Nf^2 T_F^2
\right. \nonumber \\
&& \left. ~~~
- 20977056 \zeta(3) \alpha C_A^3 
+ 6881760 \zeta(5) \alpha C_A^3
+ 3407958 \alpha C_A^3 
\right. \nonumber \\
&& \left. ~~~
+ 11259648 \zeta(3) \alpha C_A^2 \Nf T_F
- 4231296 \alpha C_A^2 \Nf T_F + 1492992 \zeta(3) \alpha C_A C_F \Nf T_F
\right. \nonumber \\
&& \left. ~~~
- 1710720 \alpha C_A C_F \Nf T_F 
- 2654208 \zeta(3) \alpha C_A \Nf^2 T_F^2
+ 778752 \alpha C_A \Nf^2 T_F^2 
\right. \nonumber \\
&& \left. ~~~
+ 39610296 \zeta(3) C_A^3
+ 15172920 \zeta(5) C_A^3 - 206477857 C_A^3 
\right. \nonumber \\
&& \left. ~~~
+ 21036672 \zeta(3) C_A^2 \Nf T_F
- 19906560 \zeta(5) C_A^2 \Nf T_F + 170325744 C_A^2 \Nf T_F
\right. \nonumber \\
&& \left. ~~~
- 80372736 \zeta(3) C_A C_F \Nf T_F 
- 29859840 \zeta(5) C_A C_F \Nf T_F
\right. \nonumber \\
&& \left. ~~~
+ 141862464 C_A C_F \Nf T_F 
- 14598144 \zeta(3) C_A \Nf^2 T_F^2
- 28289280 C_A \Nf^2 T_F^2 
\right. \nonumber \\
&& \left. ~~~
- 36827136 \zeta(3) C_F^2 \Nf T_F
+ 59719680 \zeta(5) C_F^2 \Nf T_F - 11860992 C_F^2 \Nf T_F
\right. \nonumber \\
&& \left. ~~~
+ 31850496 \zeta(3) C_F \Nf^2 T_F^2 
- 41711616 C_F \Nf^2 T_F^2
+ 512000 \Nf^3 T_F^3 \right] \frac{a^3}{746496} \nonumber \\
&& +~ O(a^4)
\end{eqnarray}
\begin{eqnarray}
C_A(a,\alpha) &=& 1 + \left[ 9 \alpha^2 C_A + 18 \alpha C_A + 97 C_A 
- 80 \Nf T_F \right] \frac{a}{36} \nonumber \\
&& +~ \left[ 810 \alpha^3 C_A^2 + 2430 \alpha^2 C_A^2 
+ 5184 \zeta(3) \alpha C_A^2 + 2817 \alpha C_A^2 - 2880 \alpha C_A \Nf T_F 
\right. \nonumber \\
&& \left. ~~~~
- 7776 \zeta(3) C_A^2 + 83105 C_A^2 - 20736 \zeta(3) C_A \Nf T_F 
- 69272 C_A \Nf T_F 
\right. \nonumber \\
&& \left. ~~~~
+ 41472 \zeta(3) C_F \Nf T_F - 47520 C_F \Nf T_F + 12800 \Nf^2 T_F^2 \right] 
\frac{a^2}{2592} \nonumber \\
&& +~ \left[ - 12636 \zeta(3) \alpha^4 C_A^3 + 17010 \zeta(5) \alpha^4 C_A^3 
+ 64638 \alpha^4 C_A^3 - 51516 \zeta(3) \alpha^3 C_A^3 
\right. \nonumber \\
&& \left. ~~~~
+ 19440 \zeta(5) \alpha^3 C_A^3 + 322947 \alpha^3 C_A^3 
+ 203148 \zeta(3) \alpha^2 C_A^3 - 8748 \zeta(4) \alpha^2 C_A^3 
\right. \nonumber \\
&& \left. ~~~~
- 374220 \zeta(5) \alpha^2 C_A^3 + 1094553 \alpha^2 C_A^3 
+ 15552 \zeta(3) \alpha^2 C_A^2 \Nf T_F 
\right. \nonumber \\
&& \left. ~~~~
- 303912 \alpha^2 C_A^2 \Nf T_F 
+ 4636764 \zeta(3) \alpha C_A^3 
- 34992 \zeta(4) \alpha C_A^3 
\right. \nonumber \\
&& \left. ~~~~
- 1341360 \zeta(5) \alpha C_A^3 
+ 1457685 \alpha C_A^3 
- 3670272 \zeta(3) \alpha C_A^2 \Nf T_F 
\right. \nonumber \\
&& \left. ~~~~
- 890064 \alpha C_A^2 \Nf T_F 
+ 746496 \zeta(3) \alpha C_A C_F \Nf T_F - 855360 \alpha C_A C_F \Nf T_F 
\right. \nonumber \\
&& \left. ~~~~
+ 663552 \zeta(3) \alpha C_A \Nf^2 T_F^2 
+ 64512 \alpha C_A \Nf^2 T_F^2 
- 7531056 \zeta(3) C_A^3 
\right. \nonumber \\
&& \left. ~~~~
- 26244 \zeta(4) C_A^3 
- 3414150 \zeta(5) C_A^3 
+ 44961125 C_A^3 - 7293888 \zeta(3) C_A^2 \Nf T_F 
\right. \nonumber \\
&& \left. ~~~~
+ 839808 \zeta(4) C_A^2 \Nf T_F 
+ 4976640 \zeta(5) C_A^2 \Nf T_F 
- 49928712 C_A^2 \Nf T_F 
\right. \nonumber \\
&& \left. ~~~~
+ 23514624 \zeta(3) C_A C_F \Nf T_F 
- 1119744 \zeta(4) C_A C_F \Nf T_F
\right. \nonumber \\
&& \left. ~~~~
+ 7464960 \zeta(5) C_A C_F \Nf T_F - 37099872 C_A C_F \Nf T_F 
+ 5971968 \zeta(3) C_A \Nf^2 T_F^2 
\right. \nonumber \\
&& \left. ~~~~
+ 13873536 C_A \Nf^2 T_F^2 + 9206784 \zeta(3) C_F^2 \Nf T_F 
- 14929920 \zeta(5) C_F^2 \Nf T_F 
\right. \nonumber \\
&& \left. ~~~~
+ 2965248 C_F^2 \Nf T_F - 12939264 \zeta(3) C_F \Nf^2 T_F^2 
+ 16130304 C_F \Nf^2 T_F^2 
\right. \nonumber \\
&& \left. ~~~~
- 1024000 \Nf^3 T_F^3 \right] \frac{a^3}{93312} ~+~ O(a^4)
\end{eqnarray}
\begin{eqnarray}
C_c(a,\alpha) &=& 1 + C_A a \nonumber \\
&& + \left[ - 36 \zeta(3) \alpha^2 C_A + 72 \alpha^2 C_A 
+ 72 \zeta(3) \alpha C_A - 21 \alpha C_A 
\right. \nonumber \\
&& \left. ~~~
- 180 \zeta(3) C_A + 1943 C_A - 760 \Nf T_F \right] \frac{C_A a^2}{192} 
\nonumber \\
&& + \left[ - 11178 \zeta(3) \alpha^3 C_A^2 - 4860 \zeta(5) \alpha^3 C_A^2 
+ 29241 \alpha^3 C_A^2 - 56862 \zeta(3) \alpha^2 C_A^2 
\right. \nonumber \\
&& \left. ~~~
+ 1458 \zeta(4) \alpha^2 C_A^2 + 34020 \zeta(5) \alpha^2 C_A^2 
+ 102789 \alpha^2 C_A^2 + 254826 \zeta(3) \alpha C_A^2 
\right. \nonumber \\
&& \left. ~~~
+ 5832 \zeta(4) \alpha C_A^2 - 63180 \zeta(5) \alpha C_A^2 - 3510 \alpha C_A^2 
- 67392 \zeta(3) \alpha C_A \Nf T_F 
\right. \nonumber \\
&& \left. ~~~
+ 42984 \alpha C_A \Nf T_F 
- 728082 \zeta(3) C_A^2 + 4374 \zeta(4) C_A^2 - 63180 \zeta(5) C_A^2 
\right. \nonumber \\
&& \left. ~~~
+ 4329412 C_A^2 - 100224 \zeta(3) C_A \Nf T_F - 139968 \zeta(4) C_A \Nf T_F 
- 2650192 C_A \Nf T_F 
\right. \nonumber \\
&& \left. ~~~
+ 684288 \zeta(3) C_F \Nf T_F 
+ 186624 \zeta(4) C_F \Nf T_F - 1165104 C_F \Nf T_F 
\right. \nonumber \\
&& \left. ~~~
+ 27648 \zeta(3) \Nf^2 T_F^2 + 330304 \Nf^2 T_F^2 \right] 
\frac{C_A a^3}{31104} ~+~ O(a^4) 
\end{eqnarray}
and 
\begin{eqnarray}
C_\psi(a,\alpha) &=& 1 - \alpha C_F a \nonumber \\
&& + \left[ - 9 \alpha^2 C_A + 8 \alpha^2 C_F + 24 \zeta(3) \alpha C_A 
- 52 \alpha C_A + 24 \zeta(3) C_A 
\right. \nonumber \\
&& \left. ~~~
- 82 C_A + 5 C_F + 28 \Nf T_F \right] \frac{C_F a^2}{8} \nonumber \\
&& + \left[ 1728 \zeta(3) \alpha^3 C_A^2 - 11880 \alpha^3 C_A^2 
- 5184 \zeta(3) \alpha^3 C_A C_F + 12312 \alpha^3 C_A C_F 
\right. \nonumber \\
&& \left. ~~~
+ 3456 \zeta(3) \alpha^3 C_F^2 - 5184 \alpha^3 C_F^2 
+ 25272 \zeta(3) \alpha^2 C_A^2 - 972 \zeta(4) \alpha^2 C_A^2 
\right. \nonumber \\
&& \left. ~~~
- 6480 \zeta(5) \alpha^2 C_A^2 
- 63747 \alpha^2 C_A^2 
- 31104 \zeta(3) \alpha^2 C_A C_F 
+ 59616 \alpha^2 C_A C_F 
\right. \nonumber \\
&& \left. ~~~
+ 181440 \zeta(3) \alpha C_A^2 
- 1944 \alpha \zeta(4) C_A^2 
- 12960 \zeta(5) \alpha C_A^2 - 358191 \alpha C_A^2 
\right. \nonumber \\
&& \left. ~~~
+ 57024 \zeta(3) \alpha C_A C_F 
- 103680 \zeta(5) \alpha C_A C_F 
+ 85536 \alpha C_A C_F 
\right. \nonumber \\
&& \left. ~~~
- 41472 \zeta(3) \alpha C_A \Nf T_F 
+ 124056 \alpha C_A \Nf T_F - 11016 \alpha C_F^2 - 28512 \alpha C_F \Nf T_F 
\right. \nonumber \\
&& \left. ~~~
+ 678024 \zeta(3) C_A^2 + 22356 \zeta(4) C_A^2 - 213840 \zeta(5) C_A^2 
- 1274056 C_A^2 
\right. \nonumber \\
&& \left. ~~~
- 228096 \zeta(3) C_A C_F - 31104 \zeta(4) C_A C_F 
+ 103680 \zeta(5) C_A C_F + 215352 C_A C_F 
\right. \nonumber \\
&& \left. ~~~
- 89856 \zeta(3) C_A \Nf T_F 
+ 760768 C_A \Nf T_F 
+ 31536 C_F^2 
- 82944 \zeta(3) C_F \Nf T_F 
\right. \nonumber \\
&& \left. ~~~
+ 68256 C_F \Nf T_F - 100480 \Nf^2 T_F^2 \right] 
\frac{C_F a^3}{5184} ~+~ O(a^4) ~. 
\end{eqnarray}
We use the convention that the variables on the right hand side are in the 
$\MSbar$ scheme. 

While \cite{6} provided the renormalization group functions for massless QCD it
is possible to deduce the quark mass anomalous dimension to four loops for an 
arbitrary colour group. This requires the conversion function for the quark
mass renormalization and the four loop $\MSbar$ anomalous dimension. The latter
has been provided in \cite{24} and \cite{25}. To deduce the former we work in 
the massless theory but renormalize the associated mass operator by inserting 
it in a quark $2$-point function at zero momentum insertion. This was the 
procedure used in the original three loop $\MSbar$ renormalization of 
\cite{28,29}. We then use the renormalization condition that there is no finite
part at the subtraction point. In this computational setup we can still use the
{\sc Mincer} algorithm, \cite{12,13}. Thus we can deduce the renormalization 
constant and hence the three loop quark mass conversion function which is 
\begin{eqnarray}
C_m(a,\alpha) &=& 1 + C_F \left[ - \alpha - 4 \right] a \nonumber \\
&& + \left[ - 18 \alpha^2 C_A + 24 \alpha^2 C_F - 84 \alpha C_A + 96 \alpha C_F
+ 432 \zeta(3) C_A - 1285 C_A 
\right. \nonumber \\
&& \left. ~~~
- 288 \zeta(3) C_F + 57 C_F + 332 \Nf T_F
\right] \frac{C_F a^2}{24} \nonumber \\
&& + \left[ - 13122 \alpha^3 C_A^2 + 15552 \alpha^3 C_A C_F 
- 7776 \alpha^3 C_F^2 + 8748 \zeta(3) \alpha^2 C_A^2 - 71685 \alpha^2 C_A^2 
\right. \nonumber \\
&& \left. ~~~
- 23328 \zeta(3) \alpha^2 C_A C_F + 89424 \alpha^2 C_A C_F 
+ 46656 \zeta(3) \alpha^2 C_F^2 
- 31104 \alpha^2 C_F^2 
\right. \nonumber \\
&& \left. ~~~
+ 103032 \zeta(3) \alpha C_A^2 - 357777 \alpha C_A^2 
- 334368 \zeta(3) \alpha C_A C_F + 573804 \alpha C_A C_F 
\right. \nonumber \\
&& \left. ~~~
- 31104 \zeta(3) \alpha C_A \Nf T_F 
+ 113400 \alpha C_A \Nf T_F 
- 46656 \zeta(3) \alpha C_F^2 - 30132 \alpha C_F^2 
\right. \nonumber \\
&& \left. ~~~
+ 62208 \zeta(3) \alpha C_F \Nf T_F - 123120 \alpha C_F \Nf T_F 
+ 3368844 \zeta(3) C_A^2 - 466560 \zeta(5) C_A^2 
\right. \nonumber \\
&& \left. ~~~
- 6720046 C_A^2 
- 2493504 \zeta(3) C_A C_F 
+ 155520 \zeta(5) C_A C_F + 2028348 C_A C_F 
\right. \nonumber \\
&& \left. ~~~
- 532224 \zeta(3) C_A \Nf T_F + 186624 \zeta(4) C_A \Nf T_F 
+ 3052384 C_A \Nf T_F 
\right. \nonumber \\
&& \left. ~~~
+ 451008 \zeta(3) C_F^2 
+ 933120 \zeta(5) C_F^2 
- 2091096 C_F^2 - 331776 \zeta(3) C_F \Nf T_F 
\right. \nonumber \\
&& \left. ~~~
- 186624 \zeta(4) C_F \Nf T_F 
+ 958176 C_F \Nf T_F - 27648 \zeta(3) \Nf^2 T_F^2 
\right. \nonumber \\
&& \left. ~~~
- 240448 \Nf^2 T_F^2 \right] \frac{C_F a^3}{7776} ~+~ O(a^4) ~. 
\end{eqnarray}
Equipped with this and the result of \cite{24,25} we find the minimal MOM quark
mass anomalous dimension is 
\begin{eqnarray}
\gamma_m^{\mbox{$\mMOMs$}}(a,\alpha) &=& -~ 3 C_F a 
+ \left[ \alpha^2 C_A - 67 C_A - 6 C_F + 8 \Nf T_F \right] \frac{C_F a^2}{4}
\nonumber \\
&& +~ \left[ -~ 3 \alpha^3 C_A^2 + 24 \alpha^3 C_A C_F 
- 54 \zeta(3) \alpha^2 C_A^2 + 411 \alpha^2 C_A^2 + 108 \alpha^2 C_A C_F 
\right. \nonumber \\
&& \left. ~~~~
-~ 48 \alpha^2 C_A \Nf T_F + 396 \zeta(3) \alpha C_A^2 + 15 \alpha C_A^2 
+ 72 \alpha C_A C_F + 48 \alpha C_A \Nf T_F 
\right. \nonumber \\
&& \left. ~~~~
+~ 5634 \zeta(3) C_A^2 - 10095 C_A^2 - 4224 \zeta(3) C_A C_F + 244 C_A C_F 
\right. \nonumber \\
&& \left. ~~~~
-~ 1152 \zeta(3) C_A \Nf T_F + 3888 C_A \Nf T_F - 3096 C_F^2 
+ 1536 \zeta(3) C_F \Nf T_F 
\right. \nonumber \\
&& \left. ~~~~
+~ 736 C_F \Nf T_F - 384 \Nf^2 T_F^2 \right] \frac{C_F a^3}{48} \nonumber \\
&& +~ \left[ - 1134 \alpha^4 C_A^3 C_F \zeta(3) 
+ 2835 \alpha^4 C_A^3 C_F \zeta(5) - 10125 \alpha^4 C_A^3 C_F 
\right. \nonumber \\
&& \left. ~~~~
+ 8586 \alpha^4 C_A^2 C_F^2 
+ 648 \alpha^4 C_A^2 C_F \Nf T_F - 2592 \alpha^4 C_A C_F^3 
\right. \nonumber \\
&& \left. ~~~~
- 35316 \zeta(3) \alpha^3 C_A^3 C_F - 1620 \zeta(5) \alpha^3 C_A^3 C_F 
+ 45522 \alpha^3 C_A^3 C_F 
\right. \nonumber \\
&& \left. ~~~~
- 15552 \zeta(3) \alpha^3 C_A^2 C_F^2 
+ 43740 \alpha^3 C_A^2 C_F^2 - 9072 \alpha^3 C_A^2 C_F \Nf T_F 
\right. \nonumber \\
&& \left. ~~~~
+ 31104 \zeta(3) \alpha^3 C_A C_F^3 - 7776 \alpha^3 C_A C_F^3 
- 485352 \zeta(3) \alpha^2 C_A^3 C_F 
\right. \nonumber \\
&& \left. ~~~~
- 28350 \zeta(5) \alpha^2 C_A^3 C_F 
+ 893025 \alpha^2 C_A^3 C_F + 167832 \zeta(3) \alpha^2 C_A^2 C_F^2 
\right. \nonumber \\
&& \left. ~~~~
+ 134784 \alpha^2 C_A^2 C_F^2 + 85536 \zeta(3) \alpha^2 C_A^2 C_F \Nf T_F 
- 307152 \alpha^2 C_A^2 C_F \Nf T_F 
\right. \nonumber \\
&& \left. ~~~~
+ 160704 \zeta(3) \alpha^2 C_A C_F^3 
+ 246888 \alpha^2 C_A C_F^3 - 82944 \zeta(3) \alpha^2 C_A C_F^2 \Nf T_F 
\right. \nonumber \\
&& \left. ~~~~
- 90720 \alpha^2 C_A C_F^2 \Nf T_F + 31104 \alpha^2 C_A C_F \Nf^2 T_F^2 
- 41472 \zeta(3) \alpha^2 C_F^3 \Nf T_F 
\right. \nonumber \\
&& \left. ~~~~
+ 799956 \zeta(3) \alpha C_A^3 C_F 
- 286740 \zeta(5) \alpha C_A^3 C_F + 295551 \alpha C_A^3 C_F 
\right. \nonumber \\
&& \left. ~~~~
- 417744 \zeta(3) \alpha C_A^2 C_F^2 + 223668 \alpha C_A^2 C_F^2 
- 658368 \zeta(3) \alpha C_A^2 C_F \Nf T_F 
\right. \nonumber \\
&& \left. ~~~~
+ 115416 \alpha C_A^2 C_F \Nf T_F 
- 824256 \zeta(3) \alpha C_A C_F^3 + 432864 \alpha C_A C_F^3 
\right. \nonumber \\
&& \left. ~~~~
+ 463104 \zeta(3) \alpha C_A C_F^2 \Nf T_F - 250560 \alpha C_A C_F^2 \Nf T_F 
\right. \nonumber \\
&& \left. ~~~~
+ 165888 \zeta(3) \alpha C_A C_F \Nf^2 T_F^2 
- 46080 \alpha C_A C_F \Nf^2 T_F^2 
\right. \nonumber \\
&& \left. ~~~~
+ 248832 \zeta(3) \alpha C_F^3 \Nf T_F 
+ 20736 \alpha C_F^3 \Nf T_F 
- 110592 \zeta(3) \alpha C_F^2 \Nf^2 T_F^2 
\right. \nonumber \\
&& \left. ~~~~
+ 27648 \alpha C_F^2 \Nf^2 T_F^2 
+ 16036470 \zeta(3) C_A^3 C_F - 6334605 \zeta(5) C_A^3 C_F 
\right. \nonumber \\
&& \left. ~~~~
- 10139319 C_A^3 C_F 
- 10029096 \zeta(3) C_A^2 C_F^2 + 3421440 \zeta(5) C_A^2 C_F^2 
\right. \nonumber \\
&& \left. ~~~~
- 2188530 C_A^2 C_F^2 - 15748128 \zeta(3) C_A^2 C_F \Nf T_F 
+ 2737152 \zeta(4) C_A^2 C_F \Nf T_F 
\right. \nonumber \\
&& \left. ~~~~
+ 4147200 \zeta(5) C_A^2 C_F \Nf T_F 
+ 8403640 C_A^2 C_F \Nf T_F + 2208384 \zeta(3) C_A C_F^3 
\right. \nonumber \\
&& \left. ~~~~
+ 6842880 \zeta(5) C_A C_F^3 - 4669704 C_A C_F^3 
+ 7091712 \zeta(3) C_A C_F^2 \Nf T_F 
\right. \nonumber \\
&& \left. ~~~~
- 2737152 \zeta(4) C_A C_F^2 \Nf T_F 
+ 1244160 \zeta(5) C_A C_F^2 \Nf T_F 
\right. \nonumber \\
&& \left. ~~~~
- 2214048 C_A C_F^2 \Nf T_F 
+ 2405376 \zeta(3) C_A C_F \Nf^2 T_F^2 
\right. \nonumber \\
&& \left. ~~~~
- 995328 \zeta(4) C_A C_F \Nf^2 T_F^2 
- 2128192 C_A C_F \Nf^2 T_F^2 
- 1741824 \zeta(3) C_F^4 
\right. \nonumber \\
&& \left. ~~~~
- 817128 C_F^4 
+ 4935168 \zeta(3) C_F^3 \Nf T_F 
- 7464960 \zeta(5) C_F^3 \Nf T_F 
\right. \nonumber \\
&& \left. ~~~~
+ 3509568 C_F^3 \Nf T_F - 1327104 \zeta(3) C_F^2 \Nf^2 T_F^2 
+ 995328 \zeta(4) C_F^2 \Nf^2 T_F^2 
\right. \nonumber \\
&& \left. ~~~~
- 605568 C_F^2 \Nf^2 T_F^2 
+ 147456 \zeta(3) C_F \Nf^3 T_F^3 
- 2048 C_F \Nf^3 T_F^3 
\right. \nonumber \\
&& \left. ~~~~
+ 1244160 \zeta(3) \frac{d^{(4)}_{FA}}{\NF} - 165888 \frac{d^{(4)}_{FA}}{\NF} 
- 2488320 \zeta(3) \frac{d^{(4)}_{FF}}{\NF} \Nf 
\right. \nonumber \\
&& \left. ~~~~
+ 331776 \frac{d^{(4)}_{FF}}{\NF} \Nf \right] \frac{a^4}{5184} ~+~ O(a^5)
\end{eqnarray} 
which involves the same rank $4$ Casimirs as the $\beta$-function. We have 
checked the three loop part by the direct evaluation of the anomalous dimension
using the minimal MOM renormalization constant. Given that we are considering
a relatively new scheme we have also renormalized the flavour non-singlet
vector current. This is important since it is a conserved physical current and
its anomalous dimension is zero to all orders in perturbation theory. This is 
true in all schemes but we have checked this explicitly to three loops by 
repeating the above quark mass operator renormalization but using the vector
current, $\bar{\psi} \gamma^\mu \psi$, instead. With the minimal MOM quark
wave function renormalization constants and isolating the Lorentz channel of
the Green's function with the inserted current corresponding to the transverse
part, we have checked that the vector current renormalization constant is unity
to three loops in the minimal MOM scheme. Thus the Slavnov-Taylor identity
has been checked to this loop order with the above renormalization.  

For more practical purposes it is useful to provide the explicit numerical
expressions for $SU(3)$. Thus we have 
\begin{eqnarray}
C_g(a,\alpha) &=& 
1 + \left[ -~ 0.375000 \alpha^2 - 0.750000 \alpha + 0.555556 \Nf - 7.041667 
\right] a \nonumber \\
&& + \left[ 0.210937 \alpha^4 - 0.562500 \alpha^3 - 0.625000 \alpha^2 \Nf 
+ 0.950346 \alpha^2 - 0.416667 \alpha \Nf 
\right. \nonumber \\
&& \left. ~~~
- 7.437954 \alpha - 0.1543210 \Nf^2 
+ 24.491186 \Nf - 163.359911 \right] a^2 \nonumber \\
&& + \left[ -~ 0.131836 \alpha^6 + 0.791016 \alpha^5 + 0.585937 \alpha^4 \Nf 
- 7.768301 \alpha^4 - 0.937500 \alpha^3 \Nf 
\right. \nonumber \\
&& \left. ~~~
- 20.064612 \alpha^3 
- 0.173611 \alpha^2 \Nf^2 - 18.480594 \alpha^2 \Nf + 8.788748 \alpha^2 
\right. \nonumber \\
&& \left. ~~~
- 2.423078 \alpha \Nf^2 + 56.307562 \alpha \Nf - 530.662942 \alpha 
+ 0.085734 \Nf^3 
\right. \nonumber \\
&& \left. ~~~
- 47.581830 \Nf^2 + 1099.935641 \Nf - 5176.895449
 \right] a^3 ~+~ O(a^4) \nonumber \\
C_A(a,\alpha) &=& 
1 + \left[ 0.750000 \alpha^2 + 1.500000 \alpha - 1.111111 \Nf 
+ 8.083333 \right] a \nonumber \\
&& + \left[ 2.812500 \alpha^3 + 8.437500 \alpha^2 - 1.666667 \alpha \Nf 
+ 31.418274 \alpha + 1.234568 \Nf^2 
\right. \nonumber \\
&& \left. ~~~
- 53.912928 \Nf + 256.103491 \right] a^2
\nonumber \\
&& + \left[ 19.411733 \alpha^4 + 81.359870 \alpha^3 - 13.754707 \alpha^2 \Nf 
+ 272.349881 \alpha^2 
\right. \nonumber \\
&& \left. ~~~
+ 6.929489 \alpha \Nf^2 - 254.788106 \alpha \Nf + 1621.114903 \alpha 
- 1.371742 \Nf^3 
\right. \nonumber \\
&& \left. ~~~
+ 171.267648 \Nf^2 
- 2601.166373 \Nf 
+ 9357.562431 \right] a^3 ~+~ O(a^4) \nonumber \\
C_c(a,\alpha) &=& 
1 + 3.000000 a + \left[ 1.346529 \alpha^2 + 3.072567 \alpha - 5.937500 \Nf 
+ 80.935770 \right] a^2 \nonumber \\
&& + \left[ 9.344565 \alpha^3 + 61.885373 \alpha^2 - 5.501305 \alpha \Nf 
+ 211.462123 \alpha + 8.765877 \Nf^2 
\right. \nonumber \\
&& \left. ~~~
- 431.804136 \Nf + 2945.691833 
\right] a^3 ~+~ O(a^4) \nonumber \\
C_\psi(a,\alpha) &=& 
1 - 1.333333 \alpha a + \left[ - 2.722222 \alpha^2 - 11.575317 \alpha 
+ 2.333333 \Nf - 25.464206 \right] a^2 \nonumber \\
&& + \left[ - 16.906900 \alpha^3 
- 72.363802 \alpha^2 + 23.739312 \alpha \Nf - 317.382214 \alpha 
\right. \nonumber \\
&& \left. ~~~
- 6.460905 \Nf^2 
+ 246.442650 \Nf - 1489.980500 \right] a^3 ~+~ O(a^4) 
\nonumber \\
C_m(a,\alpha) &=& 
1 + \left[ - 1.333333 \alpha - 5.333333 \right] a \nonumber \\
&& + \left[ - 1.222222 \alpha^2 
- 6.888889 \alpha + 9.222222 \Nf - 149.040228 \right] a^2 \nonumber \\
&& + \left[ - 11.953704 \alpha^3 - 44.682370 \alpha^2 + 14.024098 \alpha \Nf 
- 269.395219 \alpha 
\right. \nonumber \\
&& \left. ~~~
- 11.731930 \Nf^2 + 713.333651 \Nf - 5598.952656 
\right] a^3 ~+~ O(a^4) ~. 
\end{eqnarray}
Clearly it would appear that the series have large corrections at three loops.
Though that for the quark wave function is best.

\sect{$SU(\Nc)$.}

Although we have given the minimal MOM scheme results to as high a loop order
as is possible for an {\em arbitrary} colour group, it is possible to provide 
the complete set at {\em four} loops for the case of $SU(\Nc)$. This is because
the four loop $\MSbar$ anomalous dimensions of the gluon, ghost and quark are
known for this colour group for an arbitrary linear covariant gauge fixing,
\cite{24,26}. Using the electronically available data files associated with
the latter article we have extended the various three loop minimal MOM results
using the same method. This is also possible since we have the mapping for the
gauge parameter between the two schemes at three loops. Thus for the gluon we 
have 
\begin{eqnarray}
\gamma_A^{\mbox{$\mMOMs$}}(a,\alpha) &=&
[3 \alpha \Nc - 13 \Nc + 4 \Nf ] \frac{a}{6} \nonumber \\
&& + \left[ - 6 \alpha^3 \Nc^3 + 17 \alpha^2 \Nc^3 - 8 \alpha^2 \Nc^2 \Nf 
+ 17 \alpha \Nc^3 - 8 \alpha \Nc^2 \Nf - 170 \Nc^3 \right. \nonumber \\
&& \left. ~~~
+ 92 \Nc^2 \Nf - 24 \Nf \right] \frac{a^2}{24 \Nc}
\nonumber \\
&& + \left[ - 165 \alpha^4 \Nc^5 + 12 \alpha^4 \Nc^4 \Nf 
+ 54 \zeta(3) \alpha^3 \Nc^5 - 126 \alpha^3 \Nc^5 - 72 \alpha^3 \Nc^4 \Nf
\right. \nonumber \\
&& \left. ~~~
- 576 \zeta(3) \alpha^2 \Nc^5 + 1761 \alpha^2 \Nc^5 
+ 72 \zeta(3) \alpha^2 \Nc^4 \Nf - 972 \alpha^2 \Nc^4 \Nf 
\right. \nonumber \\
&& \left. ~~~
+ 216 \alpha^2 \Nc^2 \Nf - 774 \zeta(3) \alpha \Nc^5 + 102 \alpha \Nc^5 
- 144 \zeta(3) \alpha \Nc^4 \Nf - 588 \alpha \Nc^4 \Nf 
\right. \nonumber \\
&& \left. ~~~
+ 288 \alpha \Nc^2 \Nf 
+ 3456 \zeta(3) \Nc^5 - 23032 \Nc^5 - 1080 \zeta(3) \Nc^4 \Nf 
+ 15500 \Nc^4 \Nf 
\right. \nonumber \\
&& \left. ~~~
- 1360 \Nc^3 \Nf^2 + 4224 \zeta(3) \Nc^2 \Nf - 4768 \Nc^2 \Nf
- 768 \zeta(3) \Nc \Nf^2 + 736 \Nc \Nf^2 
\right. \nonumber \\
&& \left. ~~~
- 72 \Nf \right] \frac{a^3}{288 \Nc^2}
\nonumber \\
&& + \left[ 1728 \zeta(3) \alpha^5 \Nc^7 - 7560 \zeta(5) \alpha^5 \Nc^7 
- 23490 \alpha^5 \Nc^7 + 1728 \alpha^5 \Nc^6 \Nf 
\right. \nonumber \\
&& \left. ~~~
+ 52389 \zeta(3) \alpha^4 \Nc^7 - 2250 \zeta(5) \alpha^4 \Nc^7 
- 130590 \alpha^4 \Nc^7 - 1440 \zeta(3) \alpha^4 \Nc^6 \Nf 
\right. \nonumber \\
&& \left. ~~~
- 5040 \zeta(5) \alpha^4 \Nc^6 \Nf + 10440 \alpha^4 \Nc^6 \Nf 
+ 3888 \zeta(3) \alpha^4 \Nc^5 + 2430 \zeta(5) \alpha^4 \Nc^5 
\right. \nonumber \\
&& \left. ~~~
- 3888 \alpha^4 \Nc^4 \Nf + 38214 \zeta(3) \alpha^3 \Nc^7 
+ 40410 \zeta(5) \alpha^3 \Nc^7 - 20727 \alpha^3 \Nc^7 
\right. \nonumber \\
&& \left. ~~~
+ 23616 \zeta(3) \alpha^3 \Nc^6 \Nf + 1440 \zeta(5) \alpha^3 \Nc^6 \Nf 
- 86328 \alpha^3 \Nc^6 \Nf 
\right. \nonumber \\
&& \left. ~~~
+ 1944 \zeta(3) \alpha^3 \Nc^5 
+ 53460 \zeta(5) \alpha^3 \Nc^5 + 3888 \alpha^3 \Nc^4 \Nf 
- 937188 \zeta(3) \alpha^2 \Nc^7 
\right. \nonumber \\
&& \left. ~~~
+ 656370 \zeta(5) \alpha^2 \Nc^7 
+ 1387395 \alpha^2 \Nc^7 + 324144 \zeta(3) \alpha^2 \Nc^6 \Nf 
\right. \nonumber \\
&& \left. ~~~
- 30240 \zeta(5) \alpha^2 \Nc^6 \Nf 
- 1222596 \alpha^2 \Nc^6 \Nf 
- 18432 \zeta(3) \alpha^2 \Nc^5 \Nf^2 
\right. \nonumber \\
&& \left. ~~~
+ 93888 \alpha^2 \Nc^5 \Nf^2 
+ 93312 \zeta(3) \alpha^2 \Nc^5 + 145800 \zeta(5) \alpha^2 \Nc^5 
- 5832 \alpha^2 \Nc^5 
\right. \nonumber \\
&& \left. ~~~
- 319680 \zeta(3) \alpha^2 \Nc^4 \Nf 
+ 373104 \alpha^2 \Nc^4 \Nf + 55296 \zeta(3) \alpha^2 \Nc^3 \Nf^2 
\right. \nonumber \\
&& \left. ~~~
- 52992 \alpha^2 \Nc^3 \Nf^2 + 5184 \alpha^2 \Nc^2 \Nf 
- 3727014 \zeta(3) \alpha \Nc^7 
\right. \nonumber \\
&& \left. ~~~
+ 1557630 \zeta(5) \alpha \Nc^7 + 1030995 \alpha \Nc^7 
+ 1860768 \zeta(3) \alpha \Nc^6 \Nf 
\right. \nonumber \\
&& \left. ~~~
- 142560 \zeta(5) \alpha \Nc^6 \Nf - 1254696 \alpha \Nc^6 \Nf 
- 400896 \zeta(3) \alpha \Nc^5 \Nf^2 
\right. \nonumber \\
&& \left. ~~~
+ 192288 \alpha \Nc^5 \Nf^2 + 872856 \zeta(3) \alpha \Nc^5 
- 461700 \zeta(5) \alpha \Nc^5 - 23328 \alpha \Nc^5 
\right. \nonumber \\
&& \left. ~~~
+ 24576 \zeta(3) \alpha \Nc^4 \Nf^3 
- 6144 \alpha \Nc^4 \Nf^3 - 383616 \zeta(3) \alpha \Nc^4 \Nf 
\right. \nonumber \\
&& \left. ~~~
+ 428544 \alpha \Nc^4 \Nf 
+ 82944 \zeta(3) \alpha \Nc^3 \Nf^2 - 79488 \alpha \Nc^3 \Nf^2 
+ 7776 \alpha \Nc^2 \Nf 
\right. \nonumber \\
&& \left. ~~~
+ 8025711 \zeta(3) \Nc^7 + 4451400 \zeta(5) \Nc^7 - 27205691 \Nc^7 
\right. \nonumber \\
&& \left. ~~~
- 5654160 \zeta(3) \Nc^6 \Nf - 4647600 \zeta(5) \Nc^6 \Nf 
+ 23030340 \Nc^6 \Nf 
\right. \nonumber \\
&& \left. ~~~
+ 642816 \zeta(3) \Nc^5 \Nf^2 + 737280 \zeta(5) \Nc^5 \Nf^2 
- 4246944 \Nc^5 \Nf^2 
\right. \nonumber \\
&& \left. ~~~
- 2563488 \zeta(3) \Nc^5 - 4949910 \zeta(5) \Nc^5 + 142344 \Nc^5 
- 24576 \zeta(3) \Nc^4 \Nf^3 
\right. \nonumber \\
&& \left. ~~~
+ 188672 \Nc^4 \Nf^3 + 8424000 \zeta(3) \Nc^4 \Nf - 2730240 \zeta(5) \Nc^4 \Nf 
\right. \nonumber \\
&& \left. ~~~
- 7459344 \Nc^4 \Nf - 1810944 \zeta(3) \Nc^3 \Nf^2 
+ 552960 \zeta(5) \Nc^3 \Nf^2 
\right. \nonumber \\
&& \left. ~~~
+ 2377728 \Nc^3 \Nf^2 + 110592 \zeta(3) \Nc^2 \Nf^3 - 165888 \Nc^2 \Nf^3 
\right. \nonumber \\
&& \left. ~~~
- 1976832 \zeta(3) \Nc^2 \Nf + 3041280 \zeta(5) \Nc^2 \Nf - 326736 \Nc^2 \Nf 
\right. \nonumber \\
&& \left. ~~~
- 221184 \zeta(3) \Nc \Nf^2 - 552960 \zeta(5) \Nc \Nf^2 + 337536 \Nc \Nf^2 
\right. \nonumber \\
&& \left. ~~~
+ 59616 \Nf \right] \frac{a^4}{20736 \Nc^3} ~+~ O(a^5) 
\end{eqnarray}
where we have substituted the $SU(\Nc)$ values for $C_F$ and $C_A$. Similarly, 
the ghost anomalous dimension is 
\begin{eqnarray}
\gamma_c^{\mbox{$\mMOMs$}}(a,\alpha) &=&
\Nc [\alpha - 3 ] \frac{a}{4} \nonumber \\
&& + \Nc \left[ 3 \alpha^2 \Nc - 3 \alpha \Nc - 34 \Nc + 4 \Nf \right] 
\frac{a^2}{16} \nonumber \\
&& + \left[ 54 \zeta(3) \alpha^3 \Nc^3 - 45 \alpha^3 \Nc^3 
- 36 \zeta(3) \alpha^2 \Nc^3 + 216 \alpha^2 \Nc^3 - 24 \alpha^2 \Nc^2 \Nf
\right. \nonumber \\
&& \left. ~~~
+ 42 \zeta(3) \alpha \Nc^3 + 109 \alpha \Nc^3 + 48 \zeta(3) \alpha \Nc^2 \Nf 
- 76 \alpha \Nc^2 \Nf + 564 \zeta(3) \Nc^3 
\right. \nonumber \\
&& \left. ~~~
- 5196 \Nc^3 + 48 \zeta(3)\Nc^2 \Nf 
+ 1876 \Nc^2 \Nf - 160 \Nc \Nf^2 - 72 \Nf \right] \frac{a^3}{192} \nonumber \\
&& + \left[ 3843 \zeta(3) \alpha^4 \Nc^5 + 2070 \zeta(5) \alpha^4 \Nc^5 
- 8052 \alpha^4 \Nc^5 + 72 \alpha^4 \Nc^4 \Nf 
\right. \nonumber \\
&& \left. ~~~
- 3456 \zeta(3) \alpha^4 \Nc^3 + 1890 \zeta(5) \alpha^4 \Nc^3 
+ 8298 \zeta(3) \alpha^3 \Nc^5 - 6390 \zeta(5) \alpha^3 \Nc^5 
\right. \nonumber \\
&& \left. ~~~
- 9411 \alpha^3 \Nc^5 - 576 \zeta(3) \alpha^3 \Nc^4 \Nf 
+ 192 \alpha^3 \Nc^4 \Nf + 10152 \zeta(3) \alpha^3 \Nc^3 
\right. \nonumber \\
&& \left. ~~~
- 18900 \zeta(5) \alpha^3 \Nc^3 - 108 \zeta(3) \alpha^2 \Nc^5 
- 32790 \zeta(5) \alpha^2 \Nc^5 + 73071 \alpha^2 \Nc^5 
\right. \nonumber \\
&& \left. ~~~
- 2448 \zeta(3) \alpha^2 \Nc^4 \Nf + 3360 \zeta(5) \alpha^2 \Nc^4 \Nf 
- 32388 \alpha^2 \Nc^4 \Nf + 2880 \alpha^2 \Nc^3 \Nf^2 
\right. \nonumber \\
&& \left. ~~~
+ 3888 \zeta(3) \alpha^2 \Nc^3 - 22680 \zeta(5) \alpha^2 \Nc^3 
+ 648 \alpha^2 \Nc^3 + 1296 \alpha^2 \Nc^2 \Nf 
\right. \nonumber \\
&& \left. ~~~
- 40458 \zeta(3) \alpha \Nc^5 + 26790 \zeta(5) \alpha \Nc^5 
+ 88801 \alpha \Nc^5 + 1152 \zeta(3) \alpha \Nc^4 \Nf 
\right. \nonumber \\
&& \left. ~~~
- 12480 \zeta(5) \alpha \Nc^4 \Nf - 42248 \alpha \Nc^4 \Nf 
+ 3072 \zeta(3) \alpha \Nc^3 \Nf^2 + 3136 \alpha \Nc^3 \Nf^2 
\right. \nonumber \\
&& \left. ~~~
- 1080 \zeta(3) \alpha \Nc^3 - 52380 \zeta(5) \alpha \Nc^3 + 2592 \alpha \Nc^3 
- 1728 \zeta(3) \alpha \Nc^2 \Nf 
\right. \nonumber \\
&& \left. ~~~
+ 3600 \alpha \Nc^2 \Nf + 310041 \zeta(3) \Nc^5 + 192240 \zeta(5) \Nc^5 
- 1888893 \Nc^5 
\right. \nonumber \\
&& \left. ~~~
- 91728 \zeta(3) \Nc^4 \Nf - 59040 \zeta(5) \Nc^4 \Nf + 997068 \Nc^4 \Nf 
+ 13824 \zeta(3) \Nc^3 \Nf^2 
\right. \nonumber \\
&& \left. ~~~
- 141984 \Nc^3 \Nf^2 - 526176 \zeta(3) \Nc^3 + 549990 \zeta(5) \Nc^3 
+ 14904 \Nc^3 
\right. \nonumber \\
&& \left. ~~~
+ 3840 \Nc^2 \Nf^3 + 54720 \zeta(3) \Nc^2 \Nf - 34560 \zeta(5) \Nc^2 \Nf 
- 104928 \Nc^2 \Nf 
\right. \nonumber \\
&& \left. ~~~
- 4608 \zeta(3) \Nc \Nf^2 + 18816 \Nc \Nf^2 - 432 \Nf \right]
\frac{a^4}{4608 \Nc} ~+~ O(a^5) ~. 
\end{eqnarray}
Finally, the quark anomalous dimension is
\begin{eqnarray}
\gamma_\psi^{\mbox{$\mMOMs$}}(a,\alpha) &=&
\alpha [\Nc^2 - 1] \frac{a}{2 \Nc} \nonumber \\
&& + \left[ 3 \alpha^2 \Nc^4 - 3 \alpha^2 \Nc^2 + 6 \alpha \Nc^4 
- 6 \alpha \Nc^2 + 22 \Nc^4 - 4 \Nc^3 \Nf - 19 \Nc^2 
\right. \nonumber \\
&& \left. ~~~
+ 4 \Nc \Nf - 3 \right] \frac{a^2}{8 \Nc^2} \nonumber \\
&& + \left[ 18 \zeta(3) \alpha^3 \Nc^6 + 15 \alpha^3 \Nc^6 
- 18 \zeta(3) \alpha^3 \Nc^4 - 3 \alpha^3 \Nc^4 - 12 \alpha^3 \Nc^2
\right. \nonumber \\
&& \left. ~~~
- 90 \zeta(3) \alpha^2 \Nc^6 + 105 \alpha^2 \Nc^6 + 24 \alpha^2 \Nc^5 \Nf 
+ 90 \zeta(3) \alpha^2 \Nc^4 
\right. \nonumber \\
&& \left. ~~~
- 87 \alpha^2 \Nc^4 - 24 \alpha^2 \Nc^3 \Nf - 18 \alpha^2 \Nc^2 
- 618 \zeta(3) \alpha \Nc^6 + 431 \alpha \Nc^6 
\right. \nonumber \\
&& \left. ~~~
+ 96 \zeta(3) \alpha \Nc^5 \Nf - 32 \alpha \Nc^5 \Nf 
+ 618 \zeta(3) \alpha \Nc^4 - 467 \alpha \Nc^4 - 96 \zeta(3) \alpha \Nc^3 \Nf 
\right. \nonumber \\
&& \left. ~~~
+ 32 \alpha \Nc^3 \Nf + 36 \alpha \Nc^2 - 1182 \zeta(3) \Nc^6 + 3231 \Nc^6 
+ 192 \zeta(3) \Nc^5 \Nf 
\right. \nonumber \\
&& \left. ~~~
- 944 \Nc^5 \Nf + 32 \Nc^4 \Nf^2 + 894 \zeta(3) \Nc^4 - 2637 \Nc^4 
- 192 \zeta(3) \Nc^3 \Nf 
\right. \nonumber \\
&& \left. ~~~
+ 968 \Nc^3 \Nf - 32 \Nc^2 \Nf^2 + 288 \zeta(3) \Nc^2 - 576 \Nc^2 - 24 \Nc \Nf 
- 18 \right] \frac{a^3}{96 \Nc^3} \nonumber \\
&& + \left[ 330 \zeta(3) \alpha^4 \Nc^8 + 75 \zeta(5) \alpha^4 \Nc^8 
- 151 \alpha^4 \Nc^8 - 12 \alpha^4 \Nc^7 \Nf - 420 \zeta(3) \alpha^4 \Nc^6 
\right. \nonumber \\
&& \left. ~~~
- 45 \zeta(5) \alpha^4 \Nc^6 + 280 \alpha^4 \Nc^6 + 12 \alpha^4 \Nc^5 \Nf 
+ 42 \zeta(3) \alpha^4 \Nc^4 - 30 \zeta(5) \alpha^4 \Nc^4 
\right. \nonumber \\
&& \left. ~~~
- 105 \alpha^4 \Nc^4 + 48 \zeta(3) \alpha^4 \Nc^2 - 24 \alpha^4 \Nc^2 
+ 1116 \zeta(3) \alpha^3 \Nc^8 - 180 \zeta(5) \alpha^3 \Nc^8 
\right. \nonumber \\
&& \left. ~~~
- 33 \alpha^3 \Nc^8 - 192 \zeta(3) \alpha^3 \Nc^7 \Nf + 184 \alpha^3 \Nc^7 \Nf 
- 1620 \zeta(3) \alpha^3 \Nc^6 
\right. \nonumber \\
&& \left. ~~~
+ 180 \zeta(5) \alpha^3 \Nc^6 + 663 \alpha^3 \Nc^6 
+ 192 \zeta(3) \alpha^3 \Nc^5 \Nf - 184 \alpha^3 \Nc^5 \Nf 
\right. \nonumber \\
&& \left. ~~~
+ 360 \zeta(3) \alpha^3 \Nc^4 
- 558 \alpha^3 \Nc^4 + 144 \zeta(3) \alpha^3 \Nc^2 - 72 \alpha^3 \Nc^2 
- 1662 \zeta(3) \alpha^2 \Nc^8 
\right. \nonumber \\
&& \left. ~~~
- 192 \zeta(3) \Nc^3 \Nf + 3130 \zeta(5) \alpha^2 \Nc^8 - 5767 \alpha^2 \Nc^8 
- 384 \zeta(3) \alpha^2 \Nc^7 \Nf 
\right. \nonumber \\
&& \left. ~~~
- 160 \zeta(5) \alpha^2 \Nc^7 \Nf + 2580 \alpha^2 \Nc^7 \Nf 
- 96 \alpha^2 \Nc^6 \Nf^2 + 1518 \zeta(3) \alpha^2 \Nc^6 
\right. \nonumber \\
&& \left. ~~~
- 4510 \zeta(5) \alpha^2 \Nc^6 + 6307 \alpha^2 \Nc^6 
+ 576 \zeta(3) \alpha^2 \Nc^5 \Nf + 160 \zeta(5) \alpha^2 \Nc^5 \Nf 
\right. \nonumber \\
&& \left. ~~~
- 2940 \alpha^2 \Nc^5 \Nf + 96 \alpha^2 \Nc^4 \Nf^2 
+ 144 \zeta(3) \alpha^2 \Nc^4 + 1380 \zeta(5) \alpha^2 \Nc^4 
\right. \nonumber \\
&& \left. ~~~
- 558 \alpha^2 \Nc^4 - 192 \zeta(3) \alpha^2 \Nc^3 \Nf 
+ 360 \alpha^2 \Nc^3 \Nf + 18 \alpha^2 \Nc^2 
\right. \nonumber \\
&& \left. ~~~
- 33680 \zeta(3) \alpha \Nc^8 + 17420 \zeta(5) \alpha \Nc^8 
+ 14292 \alpha \Nc^8 + 7040 \zeta(3) \alpha \Nc^7 \Nf 
\right. \nonumber \\
&& \left. ~~~
- 3200 \zeta(5) \alpha \Nc^7 \Nf - 3752 \alpha \Nc^7 \Nf 
+ 320 \alpha \Nc^6 \Nf^2 + 49248 \zeta(3) \alpha \Nc^6 
\right. \nonumber \\
&& \left. ~~~
- 37500 \zeta(5) \alpha \Nc^6 - 13642 \alpha \Nc^6 
- 9408 \zeta(3) \alpha \Nc^5 \Nf + 5760 \zeta(5) \alpha \Nc^5 \Nf 
\right. \nonumber \\
&& \left. ~~~
+ 2632 \alpha \Nc^5 \Nf - 192 \alpha \Nc^4 \Nf^2 - 15568 \zeta(3) \alpha \Nc^4 
+ 20080 \zeta(5) \alpha \Nc^4 
\right. \nonumber \\
&& \left. ~~~
- 122 \alpha \Nc^4 + 2368 \zeta(3) \alpha \Nc^3 \Nf 
- 2560 \zeta(5) \alpha \Nc^3 \Nf + 1024 \alpha \Nc^3 \Nf 
\right. \nonumber \\
&& \left. ~~~
- 128 \alpha \Nc^2 \Nf^2 - 528 \alpha \Nc^2 + 96 \alpha \Nc \Nf 
- 172560 \zeta(3) \Nc^8 + 61875 \zeta(5) \Nc^8 
\right. \nonumber \\
&& \left. ~~~
+ 252104 \Nc^8 + 43392 \zeta(3) \Nc^7 \Nf - 12000 \zeta(5) \Nc^7 \Nf 
- 118572 \Nc^7 \Nf 
\right. \nonumber \\
&& \left. ~~~
- 1536 \zeta(3) \Nc^6 \Nf^2 + 15392 \Nc^6 \Nf^2 + 118002 \zeta(3) \Nc^6 
- 29085 \zeta(5) \Nc^6 
\right. \nonumber \\
&& \left. ~~~
- 225318 \Nc^6 - 640 \Nc^5 \Nf^3 - 36864 \zeta(3) \Nc^5 \Nf 
+ 480 \zeta(5) \Nc^5 \Nf 
\right. \nonumber \\
&& \left. ~~~
+ 120868 \Nc^5 \Nf + 2304 \zeta(3) \Nc^4 \Nf^2 - 16592 \Nc^4 \Nf^2 
+ 66462 \zeta(3) \Nc^4 
\right. \nonumber \\
&& \left. ~~~
- 55830 \zeta(5) \Nc^4 - 19731 \Nc^4 + 640 \Nc^3 \Nf^3 
- 6912 \zeta(3) \Nc^3 \Nf 
\right. \nonumber \\
&& \left. ~~~
+ 11520 \zeta(5) \Nc^3 \Nf 
+ 7152 \Nc^3 \Nf - 768 \zeta(3) \Nc^2 \Nf^2 + 1200 \Nc^2 \Nf^2 
\right. \nonumber \\
&& \left. ~~~
- 2304 \zeta(3) \Nc^2 + 7680 \zeta(5) \Nc^2 - 3974 \Nc^2 
+ 384 \zeta(3) \Nc \Nf - 9448 \Nc \Nf 
\right. \nonumber \\
&& \left. ~~~
- 9600 \zeta(3) + 15360 \zeta(5) - 3081 \right] 
\frac{a^4}{384 \Nc^4} ~+~ O(a^5) ~. 
\end{eqnarray}
For practical purposes it is perhaps more appropriate to provide the explicit
numerical values for all renormalization group functions at four loops for the 
$SU(3)$ colour group. Thus 
\begin{eqnarray}
\beta^{\mbox{$\mMOMs$}}(a,\alpha) &=& 
\left[ 0.666667 \Nf - 11.000000\right] a^2 \nonumber \\ 
&& +~ \left[ -~ 2.250000 \alpha^3 - \alpha^2 \Nf + 7.500000 \alpha^2 
- \alpha \Nf + 9.750000 \alpha \right. \nonumber \\
&& \left. ~~~~
+ 12.666667 \Nf - 102.000000 \right] a^3 
\nonumber \\
&& +~ \left[ 0.375000 \alpha^4 \Nf - 15.468750 \alpha^4 - 2.250000 \alpha^3 \Nf 
- 5.547924 \alpha^3
\right. \nonumber \\
&& \left. ~~~~
- 29.170372 \alpha^2 \Nf + 151.166003 \alpha^2 - 35.750000 \alpha \Nf 
+ 174.652162 \alpha 
\right. \nonumber \\
&& \left. ~~~~
- 19.383310 \Nf^2 + 625.386670 \Nf - 3040.482287 \right] a^4
\nonumber \\
&& +~ \left[ 2.250000 \alpha^5 \Nf - 114.265701 \alpha^5 + 4.816305 \alpha^4 \Nf
- 298.616620 \alpha^4 
\right. \nonumber \\
&& \left. ~~~~
- 61.925145 \alpha^3 \Nf - 37.364446 \alpha^3 
+ 43.033474 \alpha^2 \Nf^2 - 1495.393156 \alpha^2 \Nf 
\right. \nonumber \\
&& \left. ~~~~
+ 5540.175086 \alpha^2 + 3.385091 \alpha \Nf^3 - 83.916446 \alpha \Nf^2 
- 152.826933 \alpha \Nf
\right. \nonumber \\
&& \left. ~~~~
- 1111.191853 \alpha + 27.492640 \Nf^3 - 1625.402243 \Nf^2 + 24423.330550 \Nf 
\right. \nonumber \\
&& \left. ~~~~
- 100541.058601 \right] a^5 ~+~ O(a^6) \nonumber \\
\gamma_A^{\mbox{$\mMOMs$}}(a,\alpha) &=& 
\left[ 1.500000 \alpha + 0.666667 \Nf - 6.500000 \right] a \nonumber \\ 
&& +~ \left[ -~ 2.250000 \alpha^3 
- \alpha^2 \Nf + 6.375000 \alpha^2 - \alpha \Nf + 6.375000 \alpha 
\right. \nonumber \\
&& \left. ~~~~
+ 11.166667 \Nf - 63.750000 \right] a^2 \nonumber \\
&& +~ \left[ 
0.375000 \alpha^4 \Nf - 15.468750 \alpha^4 - 2.250000 \alpha^3 \Nf 
- 5.727087 \alpha^3 
\right. \nonumber \\
&& \left. ~~~~
- 26.920372 \alpha^2 \Nf + 100.182677 \alpha^2 - 22.784256 \alpha \Nf - 
77.661754 \alpha 
\right. \nonumber \\
&& \left. ~~~~
- 14.383310 \Nf^2 + 444.852414 \Nf - 1769.783563 
\right] a^3 \nonumber \\
&& +~ \left[ 2.250000 \alpha^5 \Nf - 114.265701 \alpha^5 + 3.972555 \alpha^4 \Nf
- 270.114333 \alpha^4 
\right. \nonumber \\
&& \left. ~~~~
- 72.936261 \alpha^3 \Nf + 287.225188 \alpha^3 
+ 31.783474 \alpha^2 \Nf^2 
\right. \nonumber \\
&& \left. ~~~~
- 1126.940395 \alpha^2 \Nf + 3789.309068 \alpha^2 + 3.385091 \alpha \Nf^3 
\right. \nonumber \\
&& \left. ~~~~
- 124.724674 \alpha \Nf^2 
+ 1081.645997 \alpha \Nf 
- 6926.344667 \alpha + 22.492640 \Nf^3 
\right. \nonumber \\
&& \left. ~~~~
- 1141.450868 \Nf^2 + 14846.203053 \Nf
- 54060.225189 \right] a^4 ~+~ O(a^5) \nonumber \\
\gamma_c^{\mbox{$\mMOMs$}}(a,\alpha) &=& 
\left[ 0.750000 \alpha - 2.250000 \right] a \nonumber \\ 
&& +~ \left[ 1.687500 \alpha^2 - 1.687500 \alpha 
+ 0.750000 \Nf - 19.125000 \right] a^2 \nonumber \\
&& 
+~ \left[ 2.799995 \alpha^3 - 1.125000 \alpha^2 \Nf + 24.289587 \alpha^2 - 
0.857872 \alpha \Nf 
\right. \nonumber \\
&& \left. ~~~~
+ 22.427774 \alpha - 2.500000 \Nf^2 + 90.267128 \Nf - 
635.349362 \right]  a^3 \nonumber \\
&& +~ \left[ 0.421875 \alpha^4 \Nf - 26.892596 \alpha^4 - 2.931942 \alpha^3 \Nf 
- 121.006552 \alpha^3 
\right. \nonumber \\
&& \left. ~~~~
+ 5.625000 \alpha^2 \Nf^2 - 185.757176 \alpha^2 \Nf 
+ 648.958841 \alpha^2 + 13.337341 \alpha \Nf^2 
\right. \nonumber \\
&& \left. ~~~~
- 314.266897 \alpha \Nf 
+ 1090.833574 \alpha + 2.500000 \Nf^3 - 241.975687 \Nf^2 
\right. \nonumber \\
&& \left. ~~~~
+ 4788.563749 \Nf - 23240.416706 \right] a^4 ~+~ O(a^5) \nonumber \\
\gamma_\psi^{\mbox{$\mMOMs$}}(a,\alpha) &=& 
1.333333 \alpha a + \left[ 3.000000 \alpha^2 + 6.000000 \alpha - 1.333333 \Nf + 
22.333333 \right] a^2 \nonumber \\
&& +~ \left[ 9.492589 \alpha^3 + 2.000000 \alpha^2 \Nf - 0.296280 \alpha^2 
+ 6.949789 \alpha \Nf 
\right. \nonumber \\
&& \left. ~~~~
- 78.967792 \alpha + 0.888889 \Nf^2 - 59.211534 \Nf 
+ 459.481285 \right] a^3 \nonumber \\
&& +~ \left[ -~ 0.750000 \alpha^4 \Nf + 61.650284 \alpha^4 
- 2.924683 \alpha^3 \Nf + 210.615979 \alpha^3 
\right. \nonumber \\
&& \left. ~~~~
- 2.00000 \alpha^2 \Nf^2 
+ 121.134099 \alpha^2 \Nf - 869.566016 \alpha^2 + 6.962963 \alpha \Nf^2
\right. \nonumber \\
&& \left. ~~~~
+ 77.834748 \alpha \Nf - 1553.524496 \alpha - 4.444444 \Nf^3 + 281.560058 \Nf^2
\right. \nonumber \\
&& \left. ~~~~
- 4934.050066 \Nf + 20300.851595 \right] a^4 ~+~ O(a^5) \nonumber \\
\gamma_m^{\mbox{$\mMOMs$}}(a,\alpha) &=& -~ 4.0 a 
+ \left[ \alpha^2 + 1.333333 \Nf - 69.666667 \right] a^2 \nonumber \\
&& +~ \left[ 1.916667 \alpha^3 - 2.000000 \alpha^2 \Nf + 98.522232 \alpha^2 
+ 2.000000 \alpha \Nf 
\right. \nonumber \\
&& \left. ~~~~
+ 130.753633 \alpha - 2.666667 \Nf^2 + 152.122739 \Nf 
- 1520.596003 \right] a^3 \nonumber \\
&& +~ \left[ 0.750000 \alpha^4 \Nf - 36.419738 \alpha^4 - 10.500000 \alpha^3 \Nf
+ 127.577470 \alpha^3
\right. \nonumber \\
&& \left. ~~~~
+ 6.000000 \alpha^2 \Nf^2 - 345.848075 \alpha^2 \Nf 
+ 3588.203465 \alpha^2 + 20.550022 \alpha \Nf^2 
\right. \nonumber \\
&& \left. ~~~~
- 551.791827 \alpha \Nf 
+ 5040.515124 \alpha + 5.632797 \Nf^3 - 156.909331 \Nf^2 
\right. \nonumber \\
&& \left. ~~~~
- 1073.781658 \Nf - 9337.969739 \right] a^4 ~+~ O(a^5)
\end{eqnarray}
We have checked that the Landau gauge expression for the $\beta$-function 
agrees with that of \cite{6}.

One interesting consequence of these expressions is that we can provide the
anomalous dimension of a particular dimension two operator which is
\begin{equation}
{\cal O} ~=~ \half A^a_\mu A^{a\,\mu} ~-~ \alpha \bar{c}^a c^a ~.
\end{equation}
It is known, \cite{30,31,32}, that ${\cal O}$ has a novel renormalization 
property. In the Landau gauge the anomalous dimension of ${\cal O}$ is the sum 
of the gluon and ghost anomalous dimensions. Moreover, in an arbitrary linear 
and nonlinear covariant gauge there is a simple generalization of this 
Slavnov-Taylor identity which was established in \cite{33}. This was based on 
the observation given in \cite{34} which were explicit three loop $\MSbar$ 
computations. The operator is of interest as it was an attempt to have a gluon 
mass term in the Lagrangian which while not gauge invariant is in fact BRST 
invariant, \cite{35}. It has seen renewed interest more recently, since it is 
believed to be the origin of dimension two power corrections in the running of 
an effective coupling constant in the low energy limit, \cite{8,9}. In 
\cite{26} the four loop $\MSbar$ Landau gauge result was given. However, the 
arbitrary $\alpha$ $\MSbar$ expression for a linear covariant gauge fixing was 
not recorded. As the gluon and ghost propagators are examined in the minimal 
MOM scheme, \cite{36}, it is worth providing the renormalization for the 
operators explicitly. Though for reasons of space we provide the $SU(3)$ 
expression\footnote{The full expression for $SU(\Nc)$ is given in the attached 
data file.} 
\begin{eqnarray}
\left. \gamma_{{\cal O}}^{\mbox{$\MSbars$}}(a,\alpha) \right|_{SU(3)} &=& 
\left[ 27 \alpha + 8 \Nf - 105 \right] \frac{a}{12} \nonumber \\
&& +~ \left[ 108 \alpha^2 + 567 \alpha + 548 \Nf - 4041 \right] \frac{a^2}{48}
\nonumber \\
&& +~ \left[ 12393 \alpha^3 + 4374 \zeta(3) \alpha^2 + 52488 \alpha^2 
- 22356 \alpha \Nf + 17496 \zeta(3) \alpha 
\right. \nonumber \\
&& \left. ~~~~
+ 268272 \alpha 
- 12080 \Nf^2 - 28512 \zeta(3) \Nf + 437304 \Nf + 13122 \zeta(3) 
\right. \nonumber \\
&& \left. ~~~~
- 2041389 \right] \frac{a^3}{1728} \nonumber \\
&& +~ \left[ -~ 3011499 \zeta(3) \alpha^4 + 4133430 \zeta(5) \alpha^4 
+ 8030664 \alpha^4 + 20824614 \zeta(3) \alpha^3 
\right. \nonumber \\
&& \left. ~~~~
+ 708588 \zeta(4) \alpha^3 
- 10431990 \zeta(5) \alpha^3 + 39936807 \alpha^3 
- 2939328 \zeta(3) \alpha^2 \Nf 
\right. \nonumber \\
&& \left. ~~~~
+ 314928 \zeta(4) \alpha^2 \Nf 
- 8669268 \alpha^2 \Nf + 93612348 \zeta(3) \alpha^2 
\right. \nonumber \\
&& \left. ~~~~
- 3779136 \zeta(4)\alpha^2 
- 18305190 \zeta(5) \alpha^2 + 159478227 \alpha^2 
\right. \nonumber \\
&& \left. ~~~~
+ 3359232 \zeta(3) \alpha \Nf^2 
- 2796768 \alpha \Nf^2 
- 60046272 \zeta(3) \alpha \Nf 
\right. \nonumber \\
&& \left. ~~~~
- 11337408 \zeta(4) \alpha \Nf 
- 127188144 \alpha \Nf 
+ 612180666 \zeta(3) \alpha 
\right. \nonumber \\
&& \left. ~~~~
- 43696260 \zeta(4) \alpha 
- 513923130 \zeta(5) \alpha + 1146415923 \alpha 
+ 497664 \zeta(3) \Nf^3 
\right. \nonumber \\
&& \left. ~~~~
- 846464 \Nf^3 
- 19558656 \zeta(3) \Nf^2 - 6158592 \zeta(4) \Nf^2 
- 107248896 \Nf^2 
\right. \nonumber \\
&& \left. ~~~~
- 289945440 \zeta(3) \Nf + 104451120 \zeta(4) \Nf 
+ 313061760 \zeta(5) \Nf 
\right. \nonumber \\
&& \left. ~~~~
+ 2082893580 \Nf + 72636831 \zeta(3) 
- 46766808 \zeta(4) 
- 1908463680 \zeta(5) 
\right. \nonumber \\
&& \left. ~~~~
- 7232776173 \right] \frac{a^4}{373248} ~+~ O(a^5)
\end{eqnarray}
for non-zero $\alpha$. Equipped with this then the equivalent minimal MOM
scheme expression is 
\begin{eqnarray}
\left. \gamma_{{\cal O}}^{\mbox{$\mMOMs$}}(a,\alpha) \right|_{SU(3)} &=& 
\left[ 27 \alpha + 8 \Nf - 105 \right] \frac{a}{12} \nonumber \\
&& +~ \left[ - 108 \alpha^3 - 48 \alpha^2 \Nf + 387 \alpha^2 - 48 \alpha \Nf 
+ 225 \alpha + 572 \Nf - 3978 \right]\frac{a^2}{48} \nonumber \\
&& +~ \left[ 648 \alpha^4 \Nf - 26730 \alpha^4 - 3888 \alpha^3 \Nf 
+ 21870 \zeta(3) \alpha^3 - 31347 \alpha^3 
\right. \nonumber \\
&& \left. ~~~~
+ 3888 \zeta(3) \alpha^2 \Nf 
- 53136 \alpha^2 \Nf - 102060 \zeta(3) \alpha^2 + 337770 \alpha^2 
\right. \nonumber \\
&& \left. ~~~~
- 3888 \zeta(3) \alpha \Nf - 36180 \alpha \Nf - 115182 \zeta(3) \alpha 
+ 43011 \alpha 
\right. \nonumber \\
&& \left. ~~~~
- 1536 \zeta(3) \Nf^2 
- 27328 \Nf^2 - 29088 \zeta(3) \Nf 
+ 959652 \Nf 
\right. \nonumber \\
&& \left. ~~~~
+ 696924 \zeta(3) 
- 4993812 \right] \frac{a^3}{1728} \nonumber \\
&& +~ \left[ 93312 \alpha^5 \Nf + 279936 \zeta(3) \alpha^5 
- 1224720 \zeta(5) \alpha^5 - 3805380 \alpha^5 
\right. \nonumber \\
&& \left. ~~~~
- 77760 \zeta(3) \alpha^4 \Nf 
- 272160 \zeta(5) \alpha^4 \Nf + 557928 \alpha^4 \Nf 
\right. \nonumber \\
&& \left. ~~~~
+ 11078613 \zeta(3)\alpha^4 
+ 1341360 \zeta(5) \alpha^4 - 27025488 \alpha^4 
\right. \nonumber \\
&& \left. ~~~~
+ 1135296 \zeta(3) \alpha^3 \Nf 
+ 77760 \zeta(5) \alpha^3 \Nf 
- 4591728 \alpha^3 \Nf 
\right. \nonumber \\
&& \left. ~~~~
+ 13097214 \zeta(3)\alpha^3 
+ 1319490 \zeta(5) \alpha^3 
- 10218393 \alpha^3 
\right. \nonumber \\
&& \left. ~~~~
- 221184 \zeta(3)\alpha^2 \Nf^2 
+ 1817280 \alpha^2 \Nf^2 + 14990832 \zeta(3) \alpha^2 \Nf 
\right. \nonumber \\
&& \left. ~~~~
- 816480 \zeta(5) \alpha^2 \Nf - 71613396 \alpha^2 \Nf 
- 149908644 \zeta(3) \alpha^2 
\right. \nonumber \\
&& \left. ~~~~
+ 83215350 \zeta(5) \alpha^2 
+ 277974261 \alpha^2 + 147456 \zeta(3) \alpha \Nf^3 
\right. \nonumber \\
&& \left. ~~~~
- 36864 \alpha \Nf^3 
- 6801408 \zeta(3) \alpha \Nf^2 + 3556224 \alpha \Nf^2 
\right. \nonumber \\
&& \left. ~~~~
+ 98413056 \zeta(3) \alpha \Nf 
- 10730880 \zeta(5) \alpha \Nf 
- 75346200 \alpha \Nf 
\right. \nonumber \\
&& \left. ~~~~
- 617646222 \zeta(3) \alpha 
+ 259312590 \zeta(5) \alpha + 231547167 \alpha 
\right. \nonumber \\
&& \left. ~~~~
- 73728 \zeta(3) \Nf^3 + 1125120 \Nf^3 
+ 8977920 \zeta(3) \Nf^2 
\right. \nonumber \\
&& \left. ~~~~
+ 14254080 \zeta(5) \Nf^2 
- 82945888 \Nf^2 
- 276910992 \zeta(3) \Nf 
\right. \nonumber \\
&& \left. ~~~~
- 280604160 \zeta(5) \Nf 
+ 1438122060 \Nf 
+ 1437422031 \zeta(3) 
\right. \nonumber \\
&& \left. ~~~~
+ 816720570 \zeta(5) 
- 5780555523 \right] \frac{a^4}{41472} ~+~ O(a^5)
\end{eqnarray}
for $SU(3)$.

\sect{Discussion.} 

We have provided all the renormalization group functions in QCD in the minimal
momentum subtraction scheme introduced in \cite{6}. To do this we have 
explicitly renormalized the theory and applied the renormalization prescription
given in \cite{6} to define the scheme. While \cite{6} concentrated on the 
$\beta$-function the other renormalization group functions are required for 
other problems such as the infrared structure of propagators and therefore we 
have provided that information. Currently the results are known at four loops 
for the $SU(\Nc)$ colour group and at three loops for a general group. One 
feature which differs from \cite{6} rests in the renormalization of the gauge 
parameter. In \cite{6} $\alpha$ was renormalized in the $\MSbar$ way whereas 
here we have chosen to follow a fuller approach and renormalize the gauge 
parameter according to the same criterion as all the $2$-point functions. While
this differs from \cite{6} both sets of results agree in the Landau gauge which
is the main gauge of interest for practical studies of the infrared dynamics of
the gluon and ghost.

\vspace{1cm}
\noindent
{\bf Acknowledgements.} We thank Dr A. Sternbeck for valuable discussions.

\newpage
{\begin{center}
{\bf ERRATUM}
\end{center}}
\vspace{0.4cm}
{\begin{center}
{\bf {\large Renormalization group functions of QCD in the minimal MOM scheme}}
\end{center}}
{\begin{center}
J.A. Gracey
\end{center}}
{\begin{center}
Journal of Physics A: Mathematical and Theoretical {\bf 46} (2013), 225403 1-19.
\end{center}}

\vspace{0.6cm} 
\noindent 
There was an error in the derivation of the four loop quark mass anomalous 
dimension in the minimal momentum scheme. Using the conversion function and the
four loop $\MSbar$ quark mass anomalous dimension, the four loop term of the
latter was inadvertently subtracted instead of added in applying (2.9).
Accordingly several equations need to be replaced by their correct versions. 
First, the correct version of equation (3.11) is
\begin{eqnarray}
\gamma_m^{\mbox{$\mMOMs$}}(a,\alpha) &=& -~ 3 C_F a
+ \left[ \alpha^2 C_A - 67 C_A - 6 C_F + 8 \Nf T_F \right] \frac{C_F a^2}{4}
\nonumber \\
&& +~ \left[ -~ 3 \alpha^3 C_A^2 + 24 \alpha^3 C_A C_F
- 54 \zeta(3) \alpha^2 C_A^2 + 411 \alpha^2 C_A^2 + 108 \alpha^2 C_A C_F
\right. \nonumber \\
&& \left. ~~~~
-~ 48 \alpha^2 C_A \Nf T_F + 396 \zeta(3) \alpha C_A^2 + 15 \alpha C_A^2
+ 72 \alpha C_A C_F + 48 \alpha C_A \Nf T_F
\right. \nonumber \\
&& \left. ~~~~
+~ 5634 \zeta(3) C_A^2 - 10095 C_A^2 - 4224 \zeta(3) C_A C_F + 244 C_A C_F
\right. \nonumber \\
&& \left. ~~~~
-~ 1152 \zeta(3) C_A \Nf T_F + 3888 C_A \Nf T_F - 3096 C_F^2
+ 1536 \zeta(3) C_F \Nf T_F
\right. \nonumber \\
&& \left. ~~~~
+~ 736 C_F \Nf T_F - 384 \Nf^2 T_F^2 \right] \frac{C_F a^3}{48} \nonumber \\
&& +~ \left[
- 126 \zeta(3) \alpha^4 C_A^3 C_F 
+ 315 \zeta(5) \alpha^4 C_A^3 C_F 
- 1125 \alpha^4 C_A^3 C_F  
\right. \nonumber \\
&& \left. ~~~~
+ 954 \alpha^4 C_A^2 C_F^2  
+ 72 \alpha^4 C_A^2 C_F \Nf T_F 
- 288 \alpha^4 C_A C_F^3  
\right. \nonumber \\
&& \left. ~~~~
- 3924 \zeta(3) \alpha^3 C_A^3 C_F 
- 180 \zeta(5) \alpha^3 C_A^3 C_F 
+ 5058 \alpha^3 C_A^3 C_F  
\right. \nonumber \\
&& \left. ~~~~
- 1728 \zeta(3) \alpha^3 C_A^2 C_F^2 
+ 4860 \alpha^3 C_A^2 C_F^2  
- 1008 \alpha^3 C_A^2 C_F \Nf T_F 
\right. \nonumber \\
&& \left. ~~~~
+ 3456 \zeta(3) \alpha^3 C_A C_F^3 
- 864 \alpha^3 C_A C_F^3  
- 53928 \zeta(3) \alpha^2 C_A^3 C_F 
\right. \nonumber \\
&& \left. ~~~~
- 3150 \zeta(5) \alpha^2 C_A^3 C_F 
+ 99225 \alpha^2 C_A^3 C_F  
+ 18648 \zeta(3) \alpha^2 C_A^2 C_F^2 
\right. \nonumber \\
&& \left. ~~~~
+ 14976 \alpha^2 C_A^2 C_F^2  
+ 9504 \zeta(3) \alpha^2 C_A^2 C_F \Nf T_F 
- 34128 \alpha^2 C_A^2 C_F \Nf T_F 
\right. \nonumber \\
&& \left. ~~~~
+ 17856 \zeta(3) \alpha^2 C_A C_F^3 
+ 27432 \alpha^2 C_A C_F^3  
- 9216 \zeta(3) \alpha^2 C_A C_F^2 \Nf T_F 
\right. \nonumber \\
&& \left. ~~~~
- 10080 \alpha^2 C_A C_F^2 \Nf T_F 
+ 3456 \alpha^2 C_A C_F \Nf^2 T_F^2 
- 4608 \zeta(3) \alpha^2 C_F^3 \Nf T_F 
\right. \nonumber \\
&& \left. ~~~~
+ 88884 \zeta(3) \alpha C_A^3 C_F 
- 31860 \zeta(5) \alpha C_A^3 C_F 
+ 32839 \alpha C_A^3 C_F  
\right. \nonumber \\
&& \left. ~~~~
- 46416 \zeta(3) \alpha C_A^2 C_F^2 
+ 24852 \alpha C_A^2 C_F^2  
- 73152 \zeta(3) \alpha C_A^2 C_F \Nf T_F 
\right. \nonumber \\
&& \left. ~~~~
+ 12824 \alpha C_A^2 C_F \Nf T_F 
- 91584 \zeta(3) \alpha C_A C_F^3 
+ 48096 \alpha C_A C_F^3  
\right. \nonumber \\
&& \left. ~~~~
+ 51456 \zeta(3) \alpha C_A C_F^2 \Nf T_F 
- 27840 \alpha C_A C_F^2 \Nf T_F 
\right. \nonumber \\
&& \left. ~~~~
+ 18432 \zeta(3) \alpha C_A C_F \Nf^2 T_F^2 
- 5120 \alpha C_A C_F \Nf^2 T_F^2 
+ 27648 \zeta(3) \alpha C_F^3 \Nf T_F 
\right. \nonumber \\
&& \left. ~~~~
+ 2304 \alpha C_F^3 \Nf T_F 
- 12288 \zeta(3) \alpha C_F^2 \Nf^2 T_F^2 
+ 3072 \alpha C_F^2 \Nf^2 T_F^2 
\right. \nonumber \\
&& \left. ~~~~
+ 1600326 \zeta(3) C_A^3 C_F 
- 196965 \zeta(5) C_A^3 C_F 
- 2247471 C_A^3 C_F  
\right. \nonumber \\
&& \left. ~~~~
- 939240 \zeta(3) C_A^2 C_F^2 
- 126720 \zeta(5) C_A^2 C_F^2 
+ 846270 C_A^2 C_F^2  
\right. \nonumber \\
&& \left. ~~~~
- 719136 \zeta(3) C_A^2 C_F \Nf T_F 
+ 1399224 C_A^2 C_F \Nf T_F 
- 118656 \zeta(3) C_A C_F^3 
\right. \nonumber \\
&& \left. ~~~~
+ 760320 \zeta(5) C_A C_F^3 
- 1992360 C_A C_F^3  
+ 364032 \zeta(3) C_A C_F^2 \Nf T_F 
\right. \nonumber \\
&& \left. ~~~~
+ 46080 \zeta(5) C_A C_F^2 \Nf T_F 
+ 130272 C_A C_F^2 \Nf T_F 
+ 82944 \zeta(3) C_A C_F \Nf^2 T_F^2 
\right. \nonumber \\
&& \left. ~~~~
- 255552 C_A C_F \Nf^2 T_F^2 
+ 193536 \zeta(3) C_F^4 
+ 90792 C_F^4  
- 87552 \zeta(3) C_F^3 \Nf T_F 
\right. \nonumber \\
&& \left. ~~~~
- 276480 \zeta(5) C_F^3 \Nf T_F 
+ 497472 C_F^3 \Nf T_F 
+ 36864 \zeta(3) C_F^2 \Nf^2 T_F^2 
\right. \nonumber \\
&& \left. ~~~~
- 80256 C_F^2 \Nf^2 T_F^2 
+ 9216 C_F \Nf^3 T_F^3 
- 138240 \zeta(3) \frac{d^{(4)}_{FA}}{\NF} 
+ 18432 \frac{d^{(4)}_{FA}}{\NF}
\right. \nonumber \\
&& \left. ~~~~
+ 276480 \zeta(3) \frac{d^{(4)}_{FF}}{\NF} \Nf 
- 36864 \frac{d^{(4)}_{FF}}{\NF} \Nf
\right] \frac{a^4}{576} ~+~ O(a^5) 
\end{eqnarray}
Subsequently equation (4.4) should be replaced by
\begin{eqnarray}
\gamma_m^{\mbox{$\mMOMs$}}(a,\alpha) &=& -~ 4.0 a
+ \left[ \alpha^2 + 1.333333 \Nf - 69.666667 \right] a^2 \nonumber \\
&& +~ \left[ 1.916667 \alpha^3 - 2.000000 \alpha^2 \Nf + 98.522232 \alpha^2
+ 2.000000 \alpha \Nf
\right. \nonumber \\
&& \left. ~~~~
+ 130.753633 \alpha - 2.666667 \Nf^2 + 152.122739 \Nf
- 1520.596003 \right] a^3 \nonumber \\
&& +~ \left[ 0.750000 \alpha^4 \Nf - 36.419738 \alpha^4 
- 10.500000 \alpha^3 \Nf + 127.577470 \alpha^3
\right. \nonumber \\
&& \left. ~~~~
+ 6.000000 \alpha^2 \Nf^2 - 345.848075 \alpha^2 \Nf + 3588.203465 \alpha^2 
+ 20.550022 \alpha \Nf^2 
\right. \nonumber \\
&& \left. ~~~~
- 551.791827 \alpha \Nf + 5040.515124 \alpha + 2.666667 \Nf^3 
- 298.304558 \Nf^2 
\right. \nonumber \\
&& \left. ~~~~
+ 8709.238844 \Nf - 59996.997838 \right] a^4 ~+~ O(a^5) ~.
\end{eqnarray}
The remaining results are unaffected by this change. Finally, the associated
data file has also been corrected. 

\end{document}